\documentclass[Afour,sageh,times]{sagej}

\usepackage{moreverb,url}

\usepackage[colorlinks,bookmarksopen,bookmarksnumbered,citecolor=black,urlcolor=black]{hyperref}
\usepackage{multirow}
\usepackage{caption}
\usepackage{float}
\usepackage{balance}

\def\volumeyear{2021}

\begin{document}

\runninghead{Rashid, Wei, and Wang}

\title{A Survey on Social-Physical Sensing: An Emerging Sensing Paradigm that Explores the Collective Intelligence of Humans and Machines}

\author{Md Tahmid Rashid\affilnum{1}, Na Wei\affilnum{2}, Dong Wang\affilnum{1,3}}

\affiliation{\affilnum{1}Department of Computer Science and Engineering, University of Notre Dame, Notre Dame, IN, USA\\
\affilnum{2} College of Engineering, University of Illinois Urbana-Champaign, Champaign, IL, USA\\
\affilnum{3} School of Information Sciences, University of Illinois Urbana-Champaign, Champaign, IL, USA}

\corrauth{Dong Wang, School of Information Sciences
University of Illinois Urbana-Champaign,
Champaign,
IL,
USA.}

\email{dwang24@illinois.edu}

\begin{abstract}
Propelled by the omnipresence of versatile data capture, communication, and computing technologies, physical sensing has revolutionized the avenue for decisively interpreting the real world. However, various limitations hinder physical sensing's effectiveness in critical scenarios such as disaster response and urban anomaly detection. Meanwhile, social sensing is contriving as a pervasive sensing paradigm leveraging observations from human participants equipped with portable devices and ubiquitous Internet connectivity to perceive the environment. Despite its virtues, social sensing also inherently suffers from a few drawbacks (e.g., inconsistent reliability and uncertain data provenance). Motivated by the complementary strengths of the two sensing modes, social-physical sensing (SPS) is protruding as an emerging sensing paradigm that explores the collective intelligence of humans and machines to reconstruct the ``state of the world", both physically and socially. While a good number of interesting SPS applications have been studied, several critical unsolved challenges still exist in SPS. In this paper, we provide a comprehensive survey of SPS, emphasizing its definition, key enablers, state-of-the-art applications, potential research challenges, and roadmap for future work. This paper intends to bridge the knowledge gap of existing sensing-focused survey papers by thoroughly examining the various aspects of SPS crucial for building potent SPS systems.

\end{abstract}




\maketitle

	\section{Introduction} \label{sec:intro}
With the advent of high-precision transducers in conjunction with multi-faceted communication and computation hardware, \textit{physical sensing} has matured into an avenue for accurate and agile information absorption from the real world. Broadly speaking, the term physical sensing refers to the process of leveraging hardware sensors (e.g., infrared detectors, proximity sensors, and microphones) to capture the physical world stimuli and can be predominantly classified into two variants: \textit{stationary} (e.g., surveillance cameras, digital thermostats) and \textit{mobile} (e.g., unmanned aerial vehicles (UAV), robots, satellites)~\citep{mitchell2014survey}. A few notable application domains enabled by physical sensing include: i) environmental monitoring, where arrays of sensors (e.g., temperature, pressure, and humidity sensors) are utilized to assess environmental conditions~\citep{catlett2017array}; ii) traffic surveillance, in which cameras are used to identify roadside incidents such as traffic accidents~\citep{bramberger2006distributed}; iii) industrial process monitoring, where lasers and scanners are used to coordinate manufacturing processes~\citep{chen2016industrial}; and iv) personal fitness monitoring where wearable fitness trackers assess individuals' daily physical activities~\citep{banos2014physiodroid}. 

\begin{figure*}[!ht]
    \centering
   \vspace{-0.1in}
    \includegraphics[width=17cm]{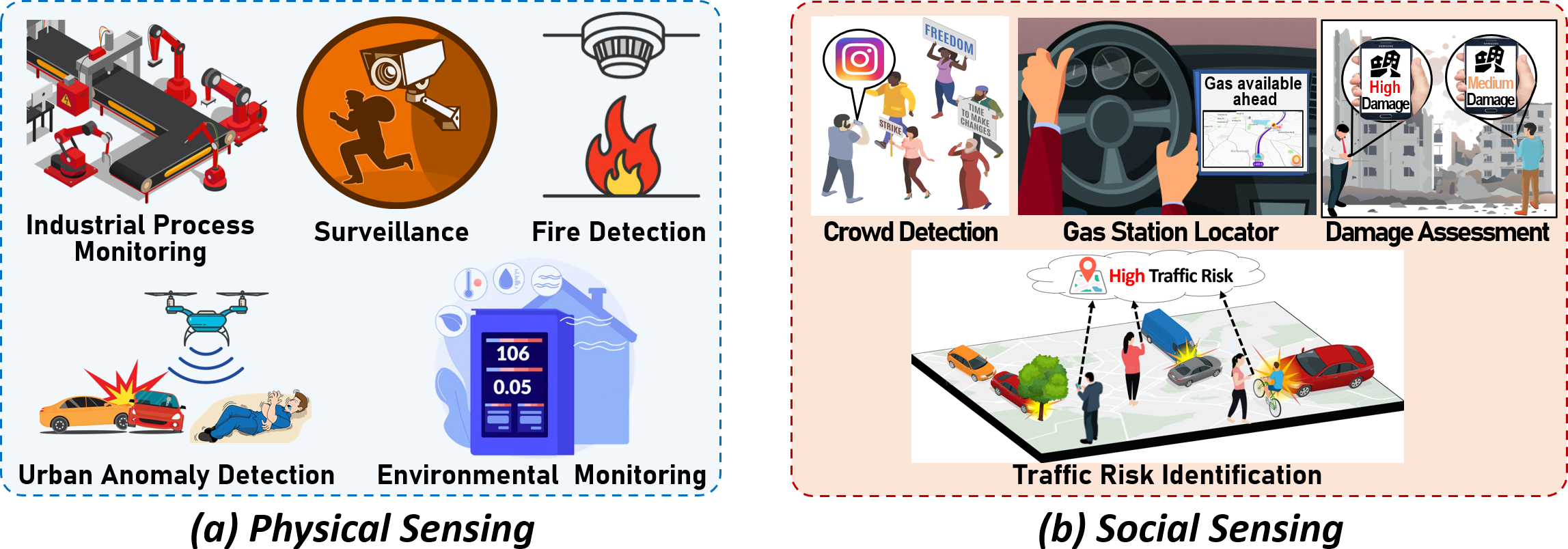}
    \captionof{figure}{(a) Examples of physical sensing applications; (b) Examples of social sensing applications}
    \vspace{-0.1in}
    \label{fig:tweetprocess}
\end{figure*}

Besides physical sensing, \textit{social sensing} has progressed as a new sensing paradigm fueled by the pervasive influence of human-centric information discovery and the widespread prevalence of Internet connectivity, where knowledge contributed by human sensors on social data collection platforms (e.g., Twitter, Waze) are acquired and analyzed to perceive real-world occurrences~\citep{wang2015social}. Social sensing can be generally categorized into two variants: \textit{social media sensing} and \textit{crowdsensing}~\citep{IPSN:14}. In social media sensing, online users proactively report occurrences around them through online social media (e.g., Twitter, Instagram, Facebook) and form virtual relationships with other users (e.g., friends or followers)~\citep{stieglitz2018social}. In crowdsensing, interested participants are assigned to carry out specialized distributed sensing tasks through various crowdsensing platforms (e.g., mobile apps such as Citizen and Waze or websites such as CrimeMapping.com). In specific scenarios, crowdsensing might be incentivized/monetized to encourage greater participation. Examples of social sensing applications include studying human mobility in urban areas~\citep{noulas2012tale}; obtaining situation awareness in the aftermath of disasters~\citep{zhang2017constraint}, poverty prediction and mapping~\citep{ledesma2020interpretable}, locating power outages in cities~\citep{hultquist2015using}, urban land usage classification~\citep{soliman2017social}, and contact tracing of contagious diseases such as COVID-19~\citep{rashid2020covidsens}. Figure~\ref{fig:tweetprocess} (a) and (b) illustrate examples of physical and social sensing applications, respectively.


While physical sensing has an established reputation for accurately capturing raw data from the environment, it suffers from several fundamental limitations such as: i) physical sensors are designed to be application-specific and are limited by the events they can sense~\citep{khalil2014wireless}, restricting their sensing scope (e.g., a temperature sensor can only capture the surrounding temperature while a microphone is designed to only record sound); ii) autonomous mobile physical sensing systems such as networks of unmanned aerial vehicles (UAVs) and unmanned ground vehicles (UGVs) do require some form of human assistance to locate events of interest, regardless of being autonomous~\citep{rashid2019collabdrone}; iii) physical sensors are typically scarce resources and need to be deployed sparingly, making their sensing coverage limited (e.g., a group of ground robots might not be able to cover a large forest during a wildfire)~\citep{casbeer2005forest}; iv) stationary physical sensors such as proximity sensors and surveillance cameras are installed in particular locations cannot be relocated easily~\citep{rashid2021unravel}; and v) physical sensors have an initial deployment cost as well as periodic maintenance costs~\citep{blaszczyszyn2008using}.

Social sensing enjoys an array of benefits not typical in physical sensing, such as: i) multifaceted information acquisition (e.g., people who report traffic incidents on social media can also report crime incidents)~\citep{wang2019age}; ii) greater mobility (e.g., human sensors tend to spontaneously move from one location to another in contrast to stationary physical sensors)~\citep{zhang2019crowdlearn}; iii) lower management costs (e.g. hardware sensors require periodic maintenance and repairs in contrast to human sensors which do not require such service from the application end)~\citep{li2019introduction}; and iv) wider sensing coverage due to the pervasive nature of social signals and the active participation of individuals (e.g., any person possessing a smart device with Internet connectivity can post on the social media from any part of the world)~\citep{IPSN:12}. However, despite its immense benefits, social sensing also has a number of drawbacks: i) inconsistent reliability since social sensing innately relies on noisy social signals contributed by unvetted human users (e.g., people can report observations that are biased or influenced by personal views)~\citep{zhang2018towards}; ii) uncertain data provenance since human sensors tend to be correlated and may propagate rumors or falsified facts initiated by other users~\citep{shang2019towards}; iii) limited sensing availability since social sensing relies on the participatory nature of individuals (e.g., people may be less interested in certain types of public occurrences and not report them through crowdsensing platforms)~\citep{zhang2018opinion}; iv) privacy concerns whereby the personal information of the participants of social sensing remains at risk of falling into the wrong hands (e.g., the whereabouts of an individual may be obtained from crowdsensing apps and used by criminals to threaten them)~\citep{pournajaf2016participant}; and v) unstructured data since human sensors can use any combination of text (which can further consist of emojis, special characters, and different languages), images, or video to report on social data platforms~\citep{zhang2016robust}.

\begin{figure}[!htb]
    \centering
    \includegraphics[width=8.5cm]{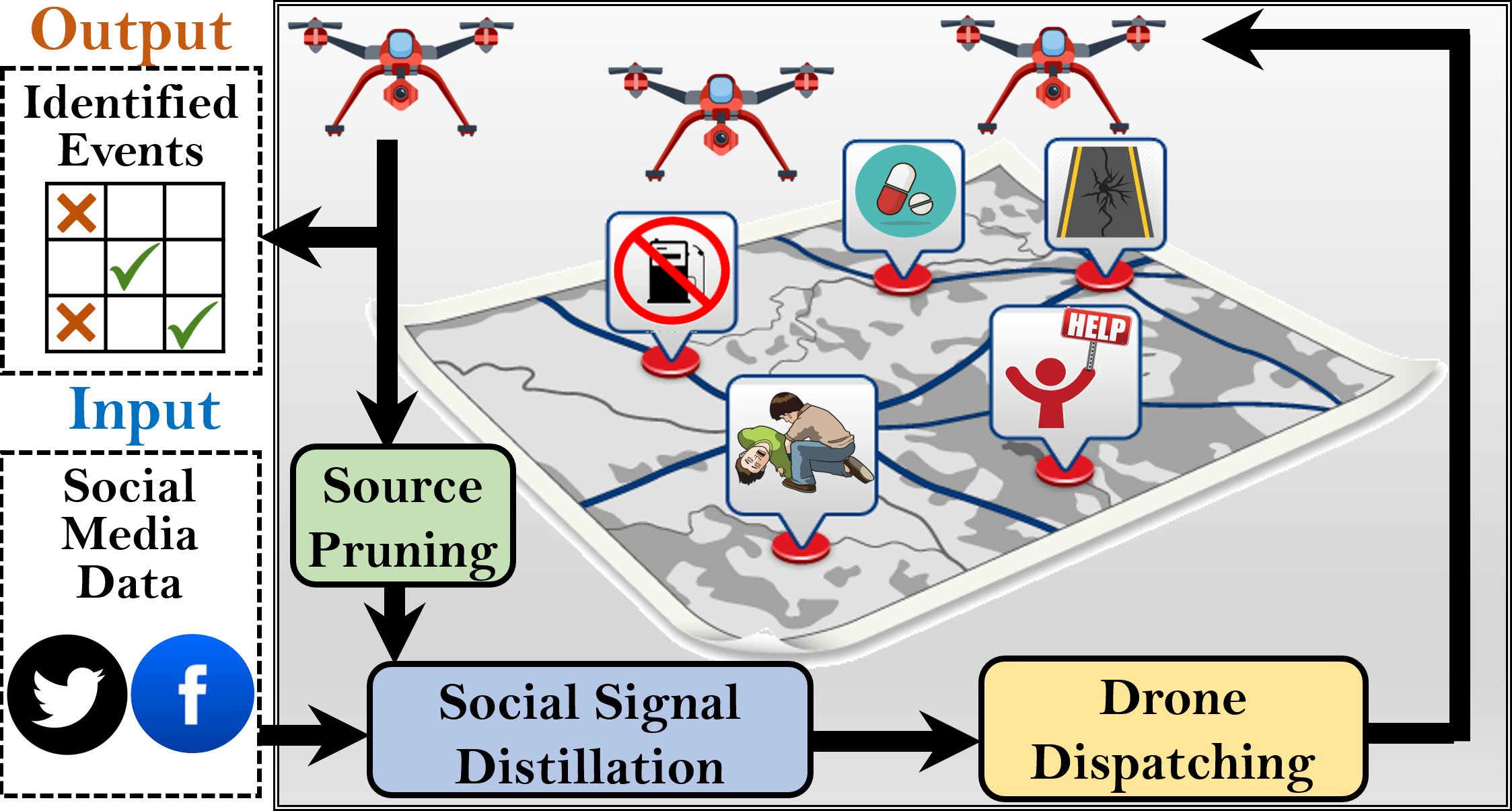}
    \caption{The architecture of a social airborne sensing (SAS) system}
    \label{fig:sas}
\end{figure}

Motivated by the complementary virtues of social and physical sensing, \textbf{social-physical sensing (SPS)} is emerging as an integrated sensing paradigm that explores the collective intelligence of both humans and machines to reconstruct the state of the world, both physically and socially~\citep{qiu2016integrating,de2017cyber,wang2013recursive}. Let us consider an SPS application known as social airborne sensing (SAS)~\citep{rashid2020socialdrone} as shown in Figure~\ref{fig:sas}. In SAS, social media signals are analyzed to discover events of interest (e.g., a building on fire) and dispatch unmanned aerial vehicles (UAVs) to validate the authenticity of the reported events using onboard sensors (e.g., cameras and thermal scanners). The validation results from the UAVs can be further used to filter out unreliable social media users. Thus SPS-based systems capitalize on the versatile sensing potentials of social and physical sensors by integrating them and mitigating their individual drawbacks for more holistic information retrieval and interpretation. In this survey paper, we explore the existing literature on SPS, emphasizing the enabling technologies behind SPS, state-of-the-art SPS applications, recurring challenges in SPS, and opportunities for future research in this emerging domain. Several recent papers on collaborative sensing, such as~\citep{chen2016industrial,o2015collaborative} present schemes that embody human discretion alongside physical sensing, exemplifying the principle of human-in-the-loop (e.g., assigning dedicated human agents to fine-tune the data captured by physical sensors). By definition, collaborative sensing leverages the cooperation of different sensors to complete large-scale sensing tasks~\citep{yi2018cyber}. While there are a few apparent similarities, a set of crucial distinctions between collaborative sensing and SPS are that: i) SPS is a much broader concept that not only considers human judgment but also explicitly models humans as ``sensors" contributing raw knowledge through social data platforms; ii) SPS applications need to characterize the dependencies between the social and physical data sources and correlate the collected data across the two sensing paradigms, a challenge which is not necessarily present for collaborative sensing applications~\citep{chen2016industrial}; and iii) human agents in collaborative sensing are often dedicated individuals~\citep{he2022collaborative} who are generally trustworthy and reliable and have high availability, whereas in SPS the human data sources can be unvetted online users on social media platforms who participate opportunistically and can be unreliable~\citep{li2019introduction}.



\begin{figure*}[!ht]
    \centering
    \includegraphics[width=17cm]{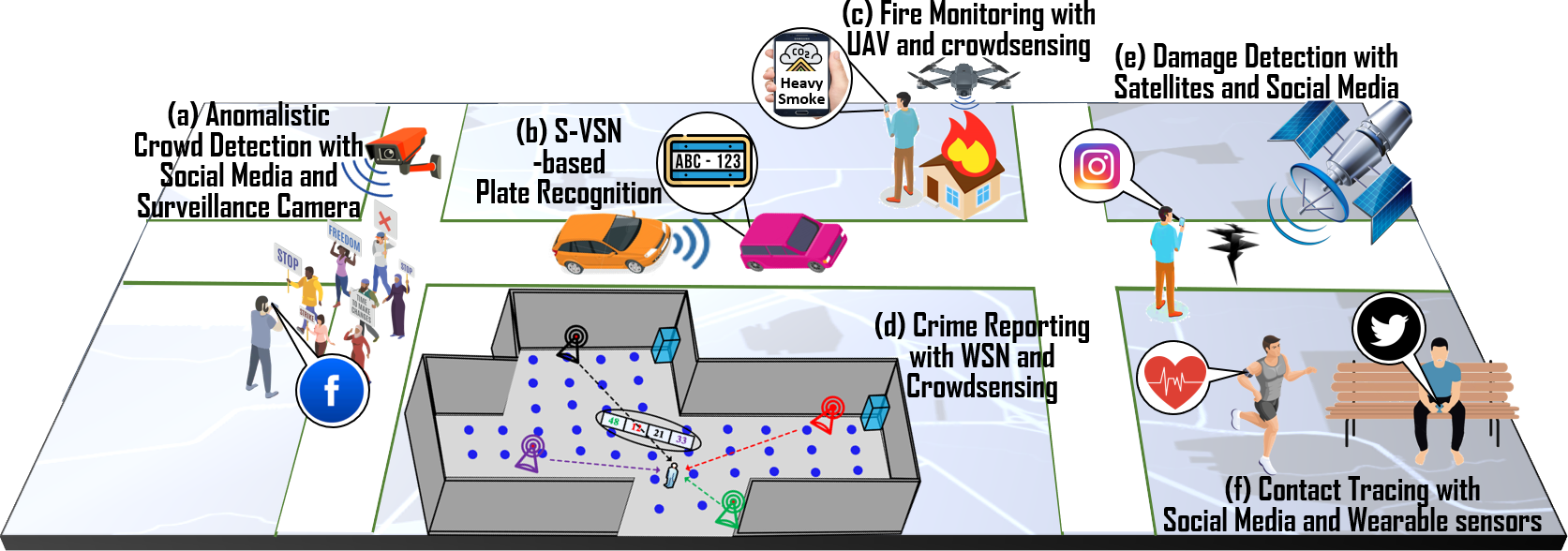}
    \caption{Examples of representative SPS applications: (a) Anomalistic Crowd Detection with Social Media and Surveillance Cameras; (b) S-VSN -based Plate Recognition; (c) Fire Monitoring with UAV and crowdsensing; (d) Crime Reporting with WSN and Crowdsensing; (e) Damage Detection with Satellites and Social Media; and (f) Contact Tracing with Social Media and Wearable sensors.}
    \label{fig:block}
\end{figure*}

A few other notable application domains empowered by SPS include urban search and rescue~\citep{dubey2019developing}, smart healthcare~\citep{chen2018urban}, simultaneous localization and mapping~\citep{jiang2019start}, human  mobility modeling~\citep{noulas2012tale}, and anomaly detection~\citep{lyu2016improved}. Figure~\ref{fig:block} highlights several recent examples of representative SPS applications which encompass: i) anomalistic crowd detection with social media and surveillance cameras; ii) social vehicular sensor network (S-VSN)-based plate recognition; iii) fire monitoring with UAV and crowdsensing; iv) road damage detection with satellites and social media; v) crime reporting with wireless sensor networks (WSN) and crowdsensing; and vi) contact tracing with social media and wearable sensors. The key design philosophy of such SPS applications is to harness the complementary information from social and physical sensors and draw a complete picture of real-world occurrences that otherwise might not be possible with standalone sensors. For instance, in an anomalistic crowd detection application based solely on networked surveillance cameras, the cameras might only be able to detect crowd events of interest (e.g., election campaigns, protests) and estimate their size without deducing the key attributes of the crowds, such as nature and cause. In contrast, people might post their plans for accumulating in public places across social media platforms (e.g., Twitter, Facebook) and post real-time updates on the progress of the crowds. However, the size and exact duration of the crowds might not be attainable from just the social media reports. When the complementary information from the social and physical sensing sources are merged, it can potentially be used to infer the critical attributes of the crowd (e.g., duration, nature, and cause of the crowd) and tell the complete story behind the crowd gathering in the first place (e.g., for staging a public demonstration in support of a protest).

While SPS promises the groundwork for a paradigm shift in sensing and data collection, it also brings new challenges to address. Examples of such challenges include: i) how to simultaneously collect relevant data from multitudes of social and physical sensors scattered around the world and relate the collected data to each other in a reliable fashion given their diverse characteristics? ii) How to efficiently handle the complex interactions between the human, cyber, and physical components in SPS when melding social sensing with physical sensing? iii) How to handle the data and device heterogeneity originating from the two distinct sensing paradigms (e.g., text data from social media vs. image data from cameras)? iv) How to characterize the dependency and correlation between the data sources when physical and social sensors are melded together? v) How to ensure end-user privacy and security considering the diverse sets of complementary information contained in the social and physical sensing mediums (e.g., geo-location data from mobile devices can be combined with information from social media posts of users to reveal sensitive information)? vi) How to adapt to the intricate dynamics that arise when jointly exploring the physical world and the social domain (e.g., how to concurrently cope with the rapidly evolving physical world events and the escalating social media reports during an emergency response)? 

Although the above challenges impose difficulty in developing effective SPS systems, they also set forth opportunities to instigate future research directions. To address the highlighted challenges, we envision the potential to incorporate techniques from multiple disciplines, such as networked sensing, communication systems, estimation theory, control theory, artificial intelligence (AI), distributed systems, and cryptography. Several current survey papers on physical sensing have investigated the functionality and features of recent physical sensing approaches (e.g., roadside surveillance systems, wildfire monitoring systems, indoor localization using wireless networks)~\citep{lee2010survey,zafari2019survey}. On the same note, several survey papers on social sensing have provided comparative studies on representative social sensing schemes (e.g., fuel availability finder using crowdsensing apps, social media-driven interesting place discovery)~\citep{ferreira2019profiling,xintong2014brief,li2016survey}. While a few survey papers have explored some sensing approaches that fall at the intersection of social sensing and physical sensing and are partially related to SPS~\citep{shi2011survey,zeng2020survey,dressler2018cyber}, they do not focus on an extensive overview of the SPS paradigm itself or present a comparative study of existing SPS applications. Most importantly, past studies have not fully addressed the need for highlighting the key challenges prevalent in emerging SPS systems, which are necessary for designing, implementing, and evaluating emerging SPS systems and applications. This survey paper aims to reduce this knowledge gap in the existing literature and extensively explore SPS.

The rest of the paper is organized as follows. Section~\ref{sec:definition} presents an in-depth overview of SPS. Section~\ref{sec:enabling} outlines the key enabling technologies for SPS. In Section~\ref{sec:app}, we identify the different applications propelled by SPS and discuss the corresponding state-of-the-art solutions. Section~\ref{sec:challenges} elucidates the key potential research challenges in constructing reliable and pervasive SPS. In Section~\ref{sec:directions}, we highlight a few research directions and opportunities for future work in SPS to mitigate the identified challenges. Lastly, in Section~\ref{sec:conclusion}, we manifest a reflection of our findings and conclude our survey of SPS.
	\section{Overview of SPS} \label{sec:definition}
This section provides a detailed overview of social-physical sensing (SPS). Specifically, we discuss the deficiency of earlier literature in defining SPS and describe the possible formats of SPS.
 
Before detailing the underpinnings of SPS, it is essential to highlight why prior studies have not acknowledged the need for a generalized definition of SPS. First, depending on the application context, the lines between social and physical sensors often tend to be blurred. For example, at first glance, an urban air quality monitoring application that uses a crowdsourcing app and social media to take user inputs for assessing the air quality might appear to be a purely social sensing application. However, if the application utilizes the GPS and accelerometers of the users' smartphones to determine the location and position of the users or relies on images taken by the users through the crowdsensing app (e.g., pictures of the sky or surroundings), the application also involves physical sensors. As such, it can be categorized as an SPS scheme. Since there are diverse ways of intertwining the plethora of social and physical sensors in applications that can be classified as SPS, there is no single widely accepted definition of SPS. Second, while SPS is a versatile sensing paradigm, it is a relatively new sensing paradigm that has not been extensively explored by existing literature. A few early survey papers have attempted to discuss sensing approaches that incorporate social and physical sensors such as cyber-physical-social systems (CPSS)~\citep{dressler2018cyber} and cyber-social systems (CSS)~\citep{wang2019data}. However, such papers solely discuss mapping physical and social sensors to cyberspace by considering the entities as black-box information retrieval tools. Moreover, survey papers on CPSS and CSS primarily focus on controlling or monitoring physical processes through feedback loops without explicitly defining SPS. 

\begin{figure}[!htb]
    \centering
   \vspace{-0.05in}
    \includegraphics[width=8.5cm]{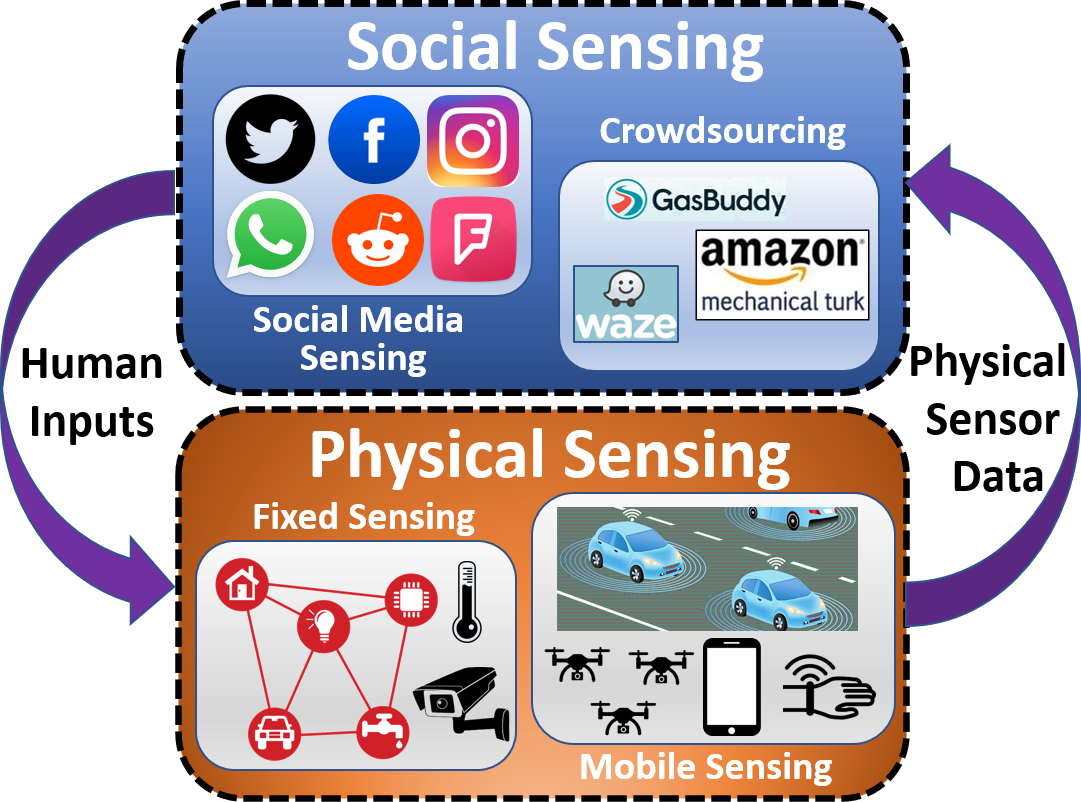}
    \caption{An overview of the SPS paradigm}
    \vspace{-0.1in}
    \label{fig:acqu}
\end{figure}

As illustrated in Figure~\ref{fig:block}, SPS encompasses several diverse domains based on the application requirements and the data acquisition tools involved. While there are no strict classification criteria for SPS schemes, the applications in SPS may be broadly classified into a few significant types, as discussed below.

The first major type of SPS involves information acquisition from reports obtained from \textbf{social media} platforms combined with sensing data from \textbf{fixed physical sensors} installed across various locations. A few examples of this form of SPS include: i) anomaly detection using surveillance cameras and social media posts~\citep{banerjee2018cyclostationary} as can be seen in Figure~\ref{fig:block} (a); and ii) traffic accident detection based on social media and roadside traffic measurement sensors~\citep{tran2018application}.

The second major type of SPS melds \textbf{social media} signals with \textbf{mobile physical sensor} data for knowledge extraction. Examples of this type of SPS are: i) contact tracing of contagious diseases such as COVID-19 with integrated social media and wearable sensors as illustrated in Figure~\ref{fig:block} (f)~\citep{rashid2022covidtrak}; ii) road damage detection using satellite imagery and social media as illustrated in Figure~\ref{fig:block} (e)~\citep{zhang2020pqa}; and iii) anomaly detection with social airborne sensing (SAS) where social media signals are used to drive UAVs to locations involved with critical events such as natural disasters as illustrated in Figure~\ref{fig:sas}~\citep{rashid2020socialdrone}.

The third major type of SPS involves \textbf{crowdsourcing} integrated with \textbf{mobile physical sensors}. A few examples of this format of SPS are: i) environmental sensors and crowdsensing-based air-quality monitoring systems~\citep{leonardi2014secondnose}; ii) noise mapping in urban areas using mobile crowdsensing and acoustic sensor networks~\citep{liu2020internet}; iii) automatic license plate recognition (ALPR) using vehicular sensors and reports from drivers on roads as shown in Figure~\ref{fig:block} (b)~\citep{zhangedgebatch}; and iv) smart water quality monitoring based on crowdsourcing and IoT-enabled water quality sensors~\citep{abualsaud2018survey}.

The fourth major type of SPS combines \textbf{crowdsourcing} with \textbf{fixed physical sensors} to perceive the environment. Some examples of this type of SPS are: i) collaborative disaster damage assessment (DDA) using surveillance camera footage and crowdsourcing website such as MTurk~\citep{zhang2019crowdlearn}; and ii) crime detection with heterogeneous sensor networks (e.g., cameras, microphones, proximity sensors) and crowdsensing apps~\citep{du2018sensable} as illustrated in Figure~\ref{fig:block} (d).

While the discussed categories represent the major formats of SPS applications, different variants of SPS can be further combined based on the application criteria since there are no absolute boundaries across the application types. For example, in a search and rescue application in the aftermath of an earthquake, locations of potential victims can be collectively gathered from social media posts and crowdsensing-based crisis reporting apps. Subsequently, ground robots might be dispatched to the reported locations to validate the information from the social data platforms.

By leveraging the collective wisdom of social and physical sensors, SPS can sense the real world and help control and actuate critical real-world processes. Examples of such control processes include mitigating traffic accidents, reducing the spread of diseases, and preventing crimes in high-risk areas. While traditional social and physical sensing systems focus on acquiring environmental stimuli, SPS applications aim to bridge the gap between the social and physical worlds by establishing a closed-loop system connecting the human, cyber, and physical worlds. To accomplish the above objectives, SPS requires careful coordination and interaction between essential enabling technologies, which are discussed in the following section.
	\section{Enabling Technologies} \label{sec:enabling}

\begin{figure*}[!htb]
    \centering
    \includegraphics[width=15cm]{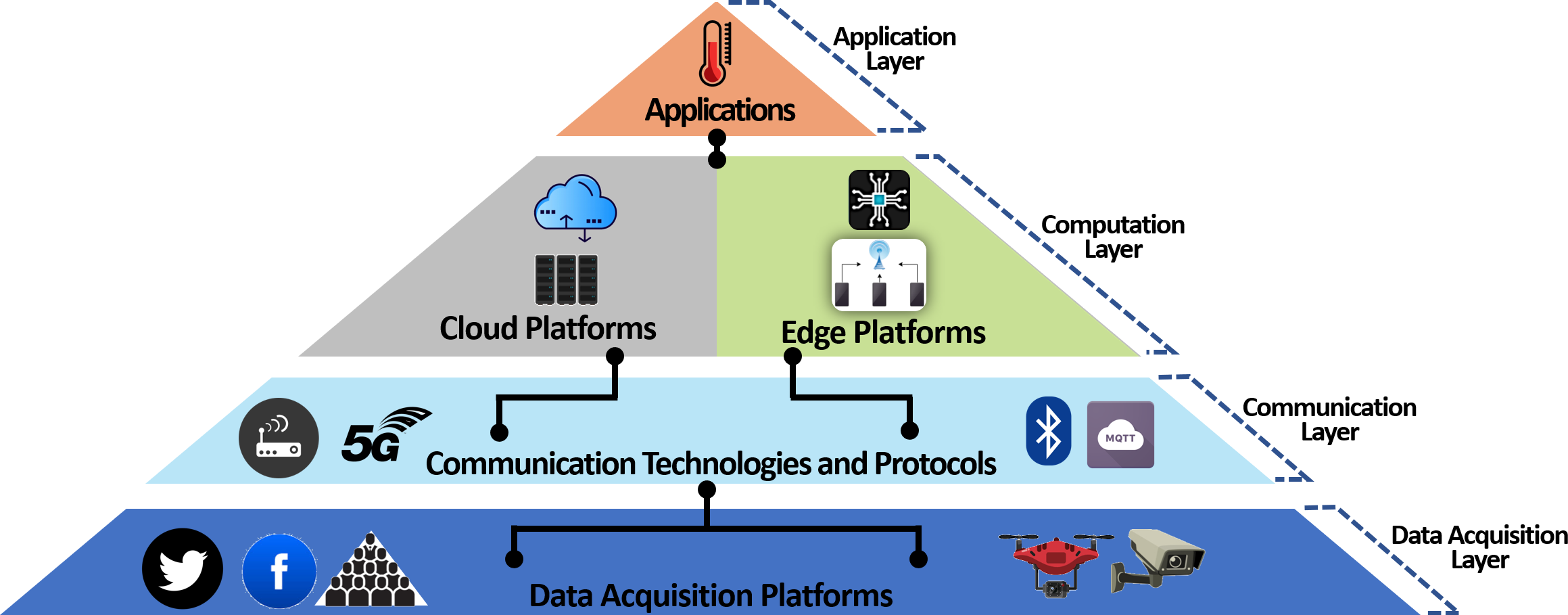}
    \caption{Abstraction layers making up SPS}
    \vspace{-0.1in}
    \label{fig:layers}
\end{figure*}

This section discusses the key enabling technologies that form the foundation of SPS. In Figure~\ref{fig:layers}, we present an abstraction model comprising the fundamental enablers for SPS. The bottom-most layer is the \textbf{data acquisition layer} containing the data acquisition tools to capture raw sensor data from the social and physical sensors in SPS (e.g., Twitter, Facebook, UAVs, and surveillance cameras). Above this layer is the \textbf{communication layer} comprising various communication technologies and protocols that enable information exchange within the entities in SPS (e.g., WiFi, 5G, Bluetooth, and MQTT). On top of the communication layer is the \textbf{computation layer}, which is further divided into \textit{cloud platforms} and \textit{edge platforms} that collectively process data in SPS. The computation layer consists of diverse processing devices (e.g., compute clusters and smartphones). At the top-most position is the \textbf{applications layer} representing the SPS applications that holistically coordinate the data acquisition, communication, and computation to capture, process, and interpret real-world phenomena. We elaborate on the applications in Section~\ref{sec:app}.

\begin{figure}[!htb]
    \centering
    \includegraphics[width=8cm]{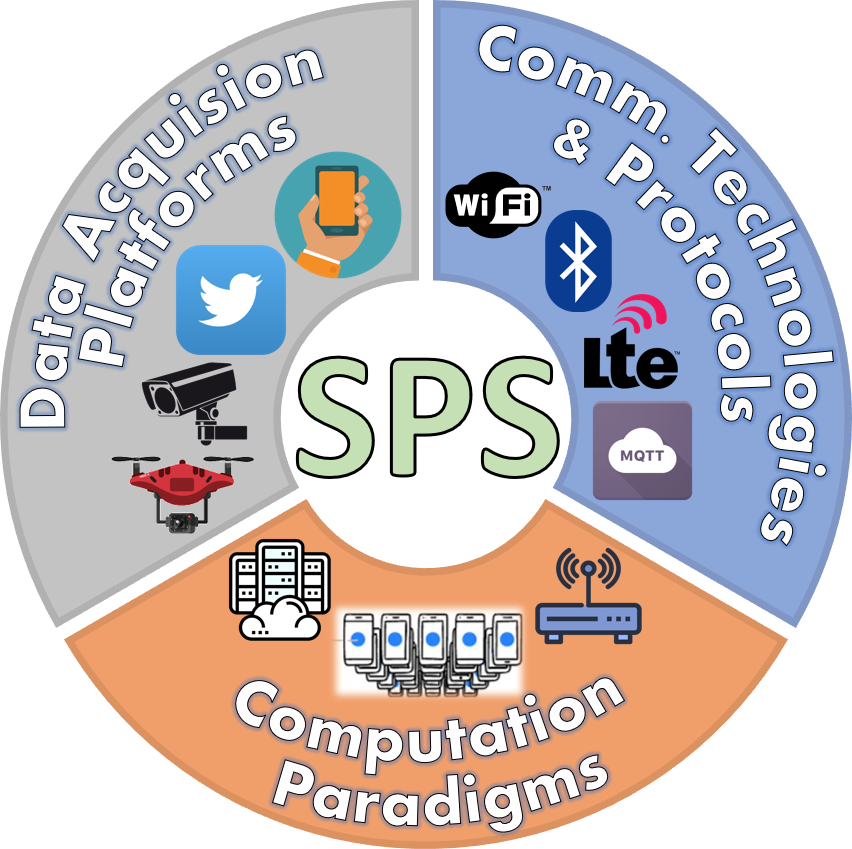}
    \caption{Examples of of the key enabling technologies for SPS: \textit{data acquisition platforms,} \textit{communication technologies and protocols,} and \textit{computing paradigms}.}
    \label{fig:enabling}
\end{figure}

Figure~\ref{fig:enabling} illustrates a few examples of the enabling technologies: i) for the data acquisition platforms, there can be any combination of sensor-fitted autonomous UAVs, surveillance cameras, social media websites like Twitter, or crowdsensing apps; ii) the communication technologies and protocols can be comprised of WiFi, Bluetooth, LTE, or MQTT; and iii) the computing paradigms can be made up of distributed compute nodes and edge devices like smartphones. In the following section, we detail the functionality of each key SPS enabler.

\subsection{Data Acquisition Platforms}
An essential component of the sensing process in SPS is data collection. The key drivers for data acquisition in SPS can be classified broadly into social and physical data platforms. The details of the platforms are discussed below.

\subsubsection{Social Data Platforms}
Intuitively, social data platforms embody the mediums of information retrieval where human sensors are directly involved in synthesizing knowledge. Recent literature such as~\citep{batrinca2015social,olteanu2019social} has extensively reviewed solutions incorporating social data platforms. Social data platforms can be further subdivided into two types.

The first type of social data platform is \textit{social media sensing} where individuals in possession of smart devices (e.g., smartphones) with Internet connectivity may voluntarily report nearby occurrences on social media portals~\citep{potts2013social,zhang2018crowdsourcing,kou2020exfaux}. Typical forms of social media include social networking services such as Twitter, Facebook, Instagram, Pinterest, and Snapchat~\citep{phua2017uses}. Within social networks, people develop connections and relationships with other individuals who are personally known to each other or who typically share similar personality traits, mutual goals, activities, ethnicity, or community~\citep{kietzmann2011social}. Conscious individuals tend to report or share incidents around them in the real world on social networking websites which serve as starting points for vital information in SPS that can be further utilized to detect the onset of critical occurrences (e.g., floods, traffic accidents, gas explosions). Another social media variant is social news aggregation websites in which news contributed by multiple individuals from different online sources is aggregated into one platform~\citep{lerman2006social}. News content in such aggregation websites is typically ranked based on popularity, credibility, and urgency. Examples of popular social news aggregation websites include Digg, Reddit, and Medium~\citep{wasike2011framing}.

The second type of social data platform is \textit{crowdsensing} which usually involves large groups of participants engaged to carry out specialized distributed sensing tasks (e.g., traffic condition reporting, crisis reporting, smart urban sensing) through individual devices (e.g., smartphones, portable sensors)~\citep{wazny2018applications}. A few representative crowdsensing applications include: i) interesting place locator~\citep{chon2012automatically}; ii) risky traffic zone identification~\citep{li2019enhancing}; and iii) urban air quality monitoring~\citep{leonardi2014secondnose}. 

Crowdsensing can be further divided into two subcategories. One variant of crowdsensing is non-monetized crowdsensing, where individuals perform small sensing tasks on a \textit{pay-it-forward} mentality with the mutual incentive of obtaining information from the platform in return. For example, in traffic apps such as Waze, drivers proactively report roadside occurrences to provide real-time traffic information in exchange for traffic updates from other users. Gas price reporting apps, such as GasBuddy, request users to report gas station availability and prices in return for providing information about gas prices at other gas stations. The other variant of crowdsensing is monetized crowdsensing, where dedicated individuals perform incentivized sensing tasks as paid freelancers. Compared to non-monetized crowdsensing, monetized crowdsensing typically attracts a more significant number of participants and is known to generate denser data~\citep{borromeo2016investigation}. Several monetized crowdsensing platforms utilize the Internet to allocate sensing tasks between participants in different parts of the world (e.g., tasks involving urban anomaly detection in a region)~\citep{singh2018safestreet}. A few examples of monetized crowdsensing applications include crisis reporting~\citep{konomi2015crowd}, gas emission monitoring in urban areas~\citep{liu2013citysee}, and health monitoring~\citep{schmitz2018leveraging}.

\subsubsection{Physical Data Platforms}
As the name implies, physical data platforms are made of hardware sensing devices for data capture (e.g., cameras and thermal scanners)~\citep{khalil2014wireless}. A good amount of effort has been contributed towards the development of energy-efficient and high-resolution transducers and electronic devices for physical sensors. Examples of such schemes can be found in~\citep{babiceanu2016big,stavropoulos2020iot}. 

The first form of physical data acquisition tools is based on \textit{fixed sensors} where a collection of dedicated sensors installed in particular locations (e.g., buildings or roadsides) are used to gather sensing data (e.g., weather sensors, infrared sensors, roadside monitoring units). The second form of physical data acquisition tools is based on \textit{mobile sensors} where the sensors are not confined to a specific location and may be transported to different locations as required. Mobile sensors can be further divided into two sub-classes. The first sub-class of mobile sensors are sensor-fitted machines such as autonomous robots, unmanned ground vehicles (UGVs), and unmanned aerial vehicles (UAVs) which are generally deployed for delay-sensitive and critical SPS applications in areas typically unreachable or dangerous to humans (e.g., locating forest fires, monitoring flood progress, searching for survivors in a wreckage site). Remote sensors are another form of mobile sensor that can obtain detailed visual representations of the constituents on the earth's surface using optical sensors installed on satellites~\citep{dash2016recent}. The second sub-class of mobile sensors utilize transducers built into smartphones (e.g., microphone, camera, and GPS), thereby eliminating the need to install or maintain dedicated sensors and providing more economical and scalable sensing compared to UAVs and UGVs. For example, the vibrations picked up by a phone's accelerometer inside a car may be utilized to locate road damage, discover potholes, or detect accidents~\citep{amin2014kalman}. Recently, there has been an emergence of mobile sensing devices such as wearable devices, health and fitness trackers, and smart tags~\citep{noulas2012tale}.

While the physical data platforms are shared with other applications, such as IoT, one crucial distinction exists. In IoT and other related applications, the data acquisition platforms only consist of fixed and mobile physical sensors~\citep{yasumoto2016survey} and often do not entail social media portals or crowdsensing apps. However, in SPS, the data sources additionally require social media and crowdsensing platforms as the fundamental drivers of knowledge. In SPS, the confluence of the social and physical data platforms helps to collect an extensive and comprehensive representation of the physical world. As an example of how the complementary information from social and physical data platforms in SPS can be leveraged to retrieve knowledge from the real world, let us consider a post-disaster resource monitoring application based on social vehicular sensor networks (S-VSN). Following a disaster (e.g., hurricane or flood), locating vital resources such as fuel and pharmacy is critical. Often people report information about such resources on social media websites such as Twitter. However, the availability of fuel at gas stations or the chances of a pharmacy being open might change at any time following the disaster. Car drivers driving nearby can be dispatched to the reported locations of the vital resources based on the tweets. Afterwards, the onboard sensors of the cars (e.g., dashboard cameras) can be used to confirm or debunk the information about the availability of the resources. Thus, the mutual information exchange between the social and physical data acquisition platforms enables SPS applications to perceive and interpret real-world phenomena with greater fidelity.

\subsection{Communication Technologies and Protocols}
The data exchange between the entities in SPS is enabled by diverse communication technologies and protocols~\citep{al2015internet}. Based on the application context (e.g., critical vs. non-critical), nature of the environment (e.g., outdoor vs. indoor), and energy profiles of the data sources (e.g., battery-powered UAVs vs fixed surveillance cameras), appropriate networking standards and protocols can be incorporated, a selection of which are discussed below.

\subsubsection{Ubiquitous Local Wireless Connectivity and Cellular Technology}
\begin{figure}[!htb]
    \centering
   \vspace{-0.1in}
    \includegraphics[width=8cm]{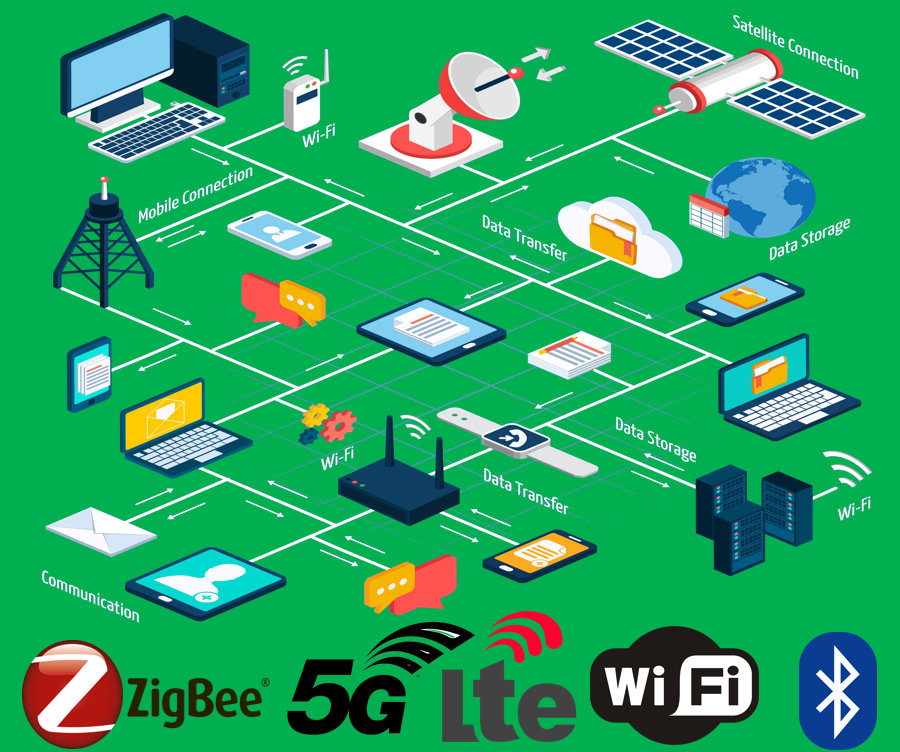}
    \caption{Overview of wireless connectivity that enables SPS}
    \label{fig:wirelesstech}
\end{figure}

In SPS, communication across the entities (e.g., UAVs, data centers, and smartphone apps) relies on ubiquitous local wireless connectivity and cellular technology. One can read more about local wireless standards and cellular technology in~\citep{mahmood2015review,sidhu2007emerging}. Commonly used connectivity methods in SPS include WiFi and Bluetooth, which utilize radio waves to transfer data among connected devices~\citep{rashid2015espionage}. For longer-range communication in SPS or fast-traveling mobile physical sensors (e.g., cars, UAVs, UGVs), cellular technology is preferred, specifically the LTE (Long-Term Evolution) and the newer 5G standards, which are treated as the norm for high-speed data transfer~\citep{sesia2011lte}. We note that the above ubiquitous local wireless connectivity and cellular technology can also be used in other related applications such as IoT as WSNs. However, in an SPS context, human sensors do not directly use such connectivity options (e.g., WiFi or LTE) to communicate their observations. Instead, human sensors leverage user interfaces (UI) on their personal devices (e.g., smartphone apps, websites on laptops) to input knowledge, eventually communicating through the highlighted ubiquitous local wireless connectivity and cellular technology. Figure~\ref{fig:wirelesstech} summarizes the state-of-the-art wireless connectivity standards enabling SPS, highlighting short-range standards such as WiFi and Bluetooth and longer-range standards such as LTE and 5G.

\subsubsection{Internet of Things (IoT) Standards and Protocols}
The interconnection of the sensing devices in bandwidth-constrained SPS applications (e.g., vehicular sensors and surveillance cameras in an anomaly detection application) is facilitated by several Internet of Things (IoT) messaging standards and protocols~\citep{al2015internet}. In the recent past, several energy-efficient IoT protocols have been developed, such as \textit{CoAP} (Constrained Application Protocol)~\citep{al2020investigating}, \textit{MQTT} (Message Queue Telemetry Transport) and \textit{XMPP} (Extensible Messaging and Presence Protocol)~\citep{al2020investigating}. While IoT deserves an elaborate discussion of its own, it is important to realize the need for IoT messaging standards that streamline communication in SPS applications. Further study about IoT applications can be found in~\citep{al2015internet}.

\subsection{Computing Paradigms}
Given the colossal amount of data generated in SPS applications, it is imperative to process and analyze the sensing signals to interpret valuable information in a scalable and efficient manner~\citep{hashem2015rise}. This paper focuses on two major computing paradigms that enable such analytics: \textit{cloud computing} and \textit{edge computing}.

\subsubsection{Cloud Computing}
Cloud computing is a distributed computing paradigm consisting of high-performance clustered computing nodes in a networked environment capable of processing huge volumes of data in parallel~\citep{qian2009cloud} and thus can serve as a powerful platform for analyzing the deluge of multi-modal data in real-time for SPS applications. Readers can find a comprehensive study of cloud applications in~\citep{rimal2009taxonomy}.

Cloud computing provides global service interfaces to the heterogeneous entities in SPS applications (e.g., vehicular sensors, smartphones, and human sensors) to upload their data which is processed using specialized hardware in conjunction with efficient task scheduling frameworks. Recent advances in cloud computing that facilitate SPS applications include: i) serverless computing, where cloud providers allocate machine resources for on-demand sensing tasks such as anomalistic crowd investigation using IoT sensors and crowdsensing~\citep{hendrickson2016serverless}; and ii) ThingSpeak, an open-source cloud framework for processing, analyzing, storing, and visualizing real-time sensing data concurrently from wearable sensors (e.g., fitness trackers and smartwatches) and social media platforms (e.g., Twitter, Facebook)~\citep{maureira2011thingspeak}.

\subsubsection{Edge Computing}
Edge computing is an efficient computing paradigm to conduct localized data processing on devices at the edge of the network~\citep{zhang2019social} and is best suited for time-critical SPS applications such as disaster response. An extensive study on edge computing-based applications can be found in~\citep{yu2017survey}. In contrast to cloud computing, edge computing administers computation at the ``edge" of the network, closer to the social and physical data sources. One key feature of edge computing is \textit{computation offloading}, where an edge device can offload data processing tasks to other idle and/or more powerful devices within a network. Delegating computation tasks from resource-constrained devices (e.g., UAVs with limited flight times) to devices with greater resource headroom (e.g., a Tesla vehicle fitted with a powerful Nvidia GPU) can speed up processing and ensure balanced resource utilization. Thus, edge computing can eliminate a single point of failure, reduce network overhead, curb transmission latency between devices, and improve response times in SPS applications.

We note that both cloud and edge computing paradigms are also incorporated in IoT and other similar applications in which they need to analyze continuous-time signals~\citep{mahmud2017signal} along with images, videos, and audio data from physical sensors~\citep{al2015internet}. However, SPS applications not only involve the above computation tasks but also require processing text data generated by human sensors, which is associated with greater computational complexity ~\citep{barkovska2021information}. Moreover, the text is often unstructured in nature and might contain misleading or sarcastic remarks that can further increase computational overhead.

The following section discusses a collection of existing representative SPS applications.

	\section{State-of-the-Art SPS Applications} \label{sec:app}

\begin{table*}[hbt!]
  \centering
  \caption{Summary and Comparison of Representative SPS Applications}
  \scalebox{1}{
	\begin{tabular}{p{2.2cm} p{2.05cm} p{3.5cm} p{7.8cm}}
\toprule
Application & Data Acquisition Platforms& Reference&Proposed Solution\\

\cmidrule(l){1-4}
    & &\citep{exposuregoogle}&Log interactions with other app users using their smartphones' Bluetooth radio and augment it with crowdsensed data\\

    &&\citep{raskar2020adding}&Extrapolate smartphone GPS data with crowdsensed data while preserving privacy to deduce approximate geographical locations of contacted persons\\
    
    \multirow{4}{6.8em}{Contact tracing of infectious diseases using crowdsensing and smartphone sensors}&\multirow{3}{9em}{Social Media + Mobile \& Fixed Physical Sensors}&\citep{bay2020bluetrace}&Exchange encrypted messages between participating devices and query suspected individuals through an app to input their contact history.\\
    
    &&\citep{altuwaiyan2018epic}&Monitor the whereabouts of infected individuals with WiFi and Bluetooth-based indoor localization and present a questionnaire through an app to input their memory of historical contacts.\\

    &&\citep{luo2020acoustic}&Analyze acoustic signals from cellular devices to measure social distance and integrate with user feedback from an app to detect infected individuals.\\

    \cmidrule(l){1-4}
    &&\citep{zhang2018risksens}&Identify locations with high traffic risk by multi-view learning from social media and satellite imagery data\\

    &&\citep{chi2017novel}&Classify land usage and land cover by melding satellite images in urban areas with localized geo-tagged social media photos.\\

    &&\citep{rosser2017rapid}&Infer flood inundation levels on different terrains by applying a Bayesian statistical model on geo-tagged images from social media, optical satellite imagery, and high-resolution terrain maps.\\

    \multirow{4}{6.8em}{Integrated social sensing and satellite-based environmental monitoring}&\multirow{3}{9em}{Social Media + Mobile Physical Sensors}&\citep{rosser2017rapid}&Detect and predict weather-driven natural disasters by fusing Twitter data with historical remote sensing data.\\

    &&\citep{zhao2020remote}&Infer socio-economic activities by converting geo-tagged tweets into high-resolution raster images and integrating them with satellite-based nighttime lights.\\
    &&\citep{huang2017cloud}&Incorporate multi-sourced data from social media, remote sensing, and online databases through spatial data mining and text mining for post-disaster damage assessment\\
    &&\citep{ghamisi2019multisource}&Combines remote sensing imagery and mobile phone positioning data for urban land usage mapping.\\

    \cmidrule(l){1-4}
    &&\citep{rashid2019collabdrone}&Identify latent correlations among reported event locations on social media to drive UAVs to regions of interest.\\

    \multirow{4}{6.8em}{Anomaly detection using SAS and S-VSN}&\multirow{4}{9em}{Social Media + Mobile Physical Sensors}&\citep{rashid2020socialdrone}&Leverage closed-loop source selection to harness the validation results from social media-driven UAVs for filtering out unreliable social media users.\\

    &&\citep{rashid2019socialcar}&Allocate incentivized sensing tasks to car drivers based on social media reports in smart city environments.\\

    &&\citep{rashid2020dasc}& Locate roads affected by damage after a disaster, such as a hurricane, and route cars avoiding damaged roads for performing sensing tasks based on social media reports.\\

    \cmidrule(l){1-4}
    
  \toprule
    \end{tabular}
    }
\label{tab:SPSsummary}
\vspace{-0.1in}
\end{table*}  
    
    \addtocounter{table}{-1}
    
    \begin{table*}[hbt!]
  \centering
  \caption{Summary and Comparison of Representative SPS Applications (Continued)}
  \scalebox{1}{
	\begin{tabular}{p{2.2cm} p{2.05cm} p{3.5cm} p{7.8cm}}
\toprule
Application & Data Acquisition Platforms& Reference&Proposed Solution\\
    \cmidrule(l){1-4}
    &&\citep{zhang2019edgebatch}& Combine reports about license plates of probable suspects from concerned citizens in crowdsensing apps with inputs from IoT sensors (e.g., surveillance cameras) to detect the license plates.\\

    \multirow{4}{6.8em}{License plate recognition using crowdsensing and physical sensors}&\multirow{3}{9em}{Crowdsensing + Fixed \& Mobile Physical Sensors}&\citep{trottier2014crowdsourcing}&Perform image processing on dashboard camera footage and combine with crowdsensed feedback to recognize the number plates.\\

    &&\citep{alcaide2014privacy} & Obtain privacy-preserved anonymous inputs from crowdsensing participants and integrate with image processing techniques to locate suspects' number plates.\\

    &&\citep{yan2011crowdpark}& Mask and protect the identity of the owners of license plates recognized using data from crowdsensing apps and roadside monitors.\\

\cmidrule(l){1-4}
     &&\citep{jun2013social}&Integrate physical traces of an individual posted through social media with RSSI signals from WiFi routers to derive their location inside a building.\\

    &&\citep{chu2020sbot}&Combines user statuses and updates posted through social media using text mining techniques with telemetry data from smartphone sensors to pinpoint users' location.\\

    &&\citep{liu2010indoor} & Perform indoor localization and visualization of complex environments such as staircases or corridors by using backpacks equipped with 2D laser scanners and inertial measurement units augmented with historical social network traces of users. \\

    \multirow{4}{6.8em}{Situational awareness using social media and crowdsensing melded with IoT (Social/CrowdIoT)}&\multirow{3}{9em}{Crowdsensing + Mobile \& Fixed Physical Sensors}&\citep{hamza2020privacy}&Geo-locate users indoors using privacy-preserving approaches to protect their identities.\\

    &&\citep{dunphy2015crowdsourcing}& Process frames from CCTV surveillance footage using AI and combine with perception from Amazon MTurk participants to tag instances of abnormal occurrences in real-time. \\

    &&\citep{abu2016enhancing}& Predict the possibility of a crisis in smart cities using crowdsensing apps and fixed urban IoT sensors (e.g., proximity sensors, acoustic sensors).\\

    &&\citep{horita2018determining}&Infer probable locations with a flood by integrating crowdsourcing data with data from \textit{in situ} weather radars.\\

    &&\citep{han2019harnessing}&Provide rapid disaster response by using vital metrics derived from both crowdsensing apps and portable devices equipped with RFID technology.\\

\toprule
    \end{tabular}
    }
\label{tab:SPSsummary2}
\vspace{-0.1in}
\end{table*}

This section reviews a few exciting real-world SPS applications from the current literature. In Table~\ref{tab:SPSsummary}, we provide a comprehensive summary of the representative SPS applications and the associated solutions. In particular, the first column of the table indicates the SPS application type, which can encompass a wide variety of areas such as healthcare, environmental monitoring, anomaly detection, license plate recognition, and situational awareness. The second column indicates the data acquisition platforms involved, which can be any combination of social and physical sensors. The third column indicates references to schemes from current literature for the particular application scenario, a brief description of which is given in the fourth column. We further detail each application scenario and its corresponding schemes in the following subsections. 

\subsection{Contact tracing of infectious diseases using crowdsensing and smartphone sensors}
\begin{figure}[!htb]
    \centering
    \includegraphics[width=8.5cm]{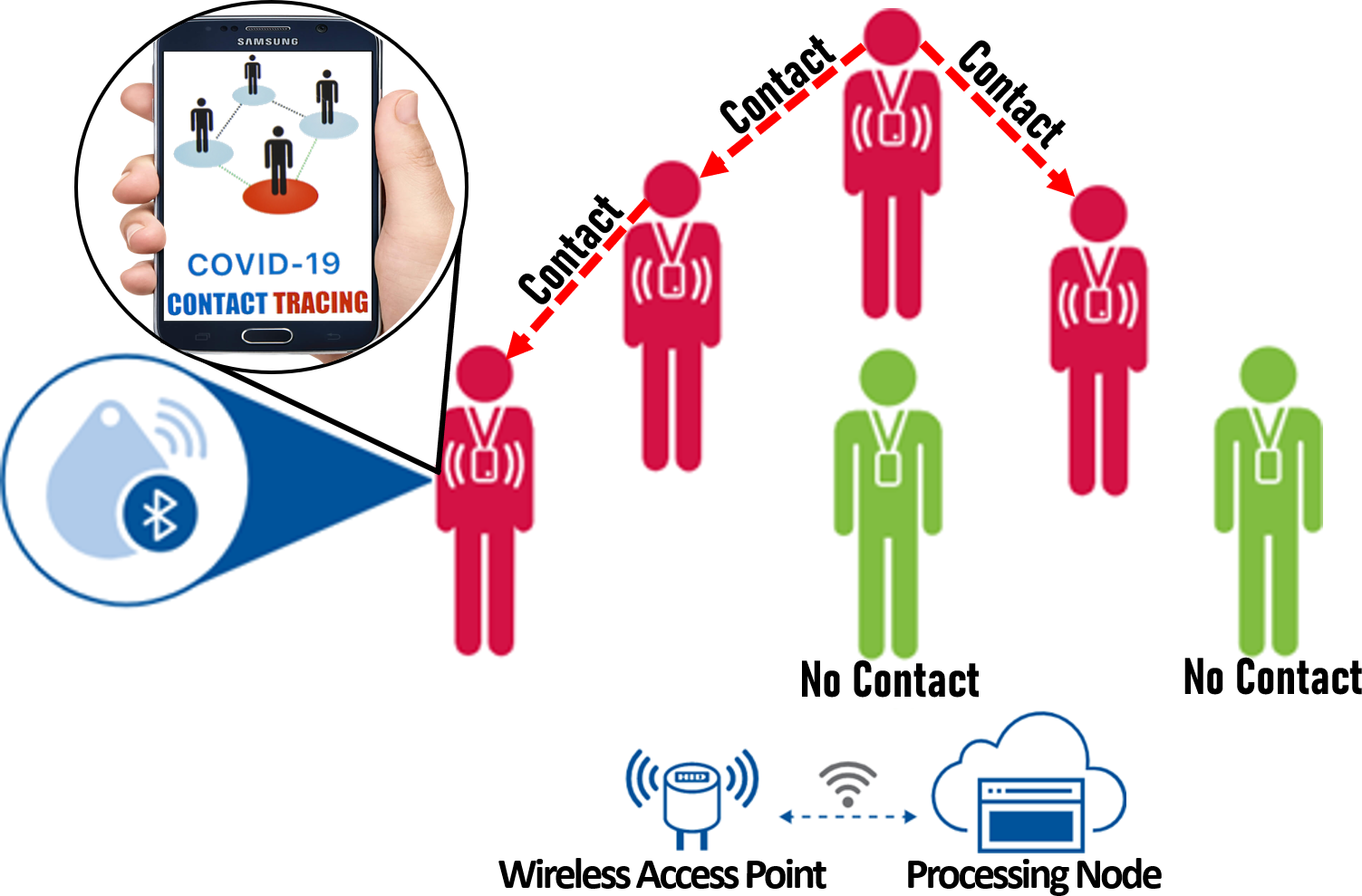}
    \caption{Concept of contact tracing with crowdsensing and smartphone sensors}
    \label{fig:contact}
\end{figure}

In the field of epidemiology, \textit{contact tracing} is a mechanism of identifying and monitoring individuals who may have come in close contact with people having any infectious disease to circumvent further disease spread~\citep{eames2003contact}. Pinpointing and quarantining sources of an infectious disease restricts their ability to ``contact" the disease, thereby minimizing community spread~\citep{altuwaiyan2018epic}. Recently, with the pandemic of the coronavirus disease 2019 (COVID-19), there has been a surge of contact tracing applications that combine the power of crowdsensing with smartphone sensors distributed around the world to study the physical footprints of users~\citep{altuwaiyan2018epic,exposuregoogle,michael2020getting,raskar2020adding,pandurangagovernment,bay2020bluetrace}. Figure~\ref{fig:contact} presents the concept of contact tracing based on crowdsensing and smartphone sensors~\citep{george2020issue}. When any individual tests positive for COVID-19 and reports his illness through a contact tracing app installed on his smartphone, his physical footprints from the GPS data on his smartphone can be analyzed to examine his whereabouts and physical encounters with other individuals. If it is found that an untested or uninfected individual came in close contact with this infected person, that particular individual can be alerted to get tested and quarantined to reduce the likelihood of further spread.

Several recent studies have attempted to meld non-monetized crowdsensing with Bluetooth and WiFi radios found in smartphones for COVID-19 contact tracing applications~\citep{pandurangagovernment,altuwaiyan2018epic,bay2020bluetrace}. For example, Google and Apple launched a decentralized COVID-19 contact tracing framework called Exposure Notification System (ENS) that logs interactions with other ENS users using their smartphones' Bluetooth radio~\citep{exposuregoogle} and augments it with crowdsensed data provided through mobile apps~\citep{michael2020getting}. MIT Media Lab further enhanced the ENS framework by developing a privacy-preserving location extrapolation mechanism with a smartphone's GPS to deduce the approximate geographical location of a contacted person~\citep{raskar2020adding}. The scheme also allows healthy users to determine if they have ``crossed paths" with any infected person~\citep{pandurangagovernment}. 

The Singaporean government launched BlueTrace, a privacy-aware open-source COVID-19 contact tracing application based on Bluetooth-based localization and voluntary crowdsensing application~\citep{bay2020bluetrace} that logs Bluetooth interactions between participating devices. When two devices ``meet", they trade encrypted messages with temporary identifiers, and anyone suspected of infection will be requested to share their contact history with the concerned authority. Altuwaiyan \textit{et al.} proposed a contact tracing scheme with integrated WiFi and Bluetooth-based localization technology from smartphones combined with crowdsensing through a mobile app~\citep{altuwaiyan2018epic}. Once users are tested positive, they are presented with a questionnaire through the app to input their memory of historical contacts. A contact tracing project called A-Turf was undertaken to accurately detect ``encounters" between users within close proximity (e.g. less than six feet) using user feedback reported through a crowdsensing app and acoustic signals emitted by smartphones~\citep{luo2020acoustic}. By determining the ``footprint" of infected individuals, crowdsensing and smartphone sensor-driven contact tracing systems help to test, isolate, and treat potential contacts of infected people.

\subsection{Integrated social sensing and satellite-based environmental monitoring}
\begin{figure}[!htb]
    \centering
   \vspace{-0.1in}
    \includegraphics[width=8.5cm]{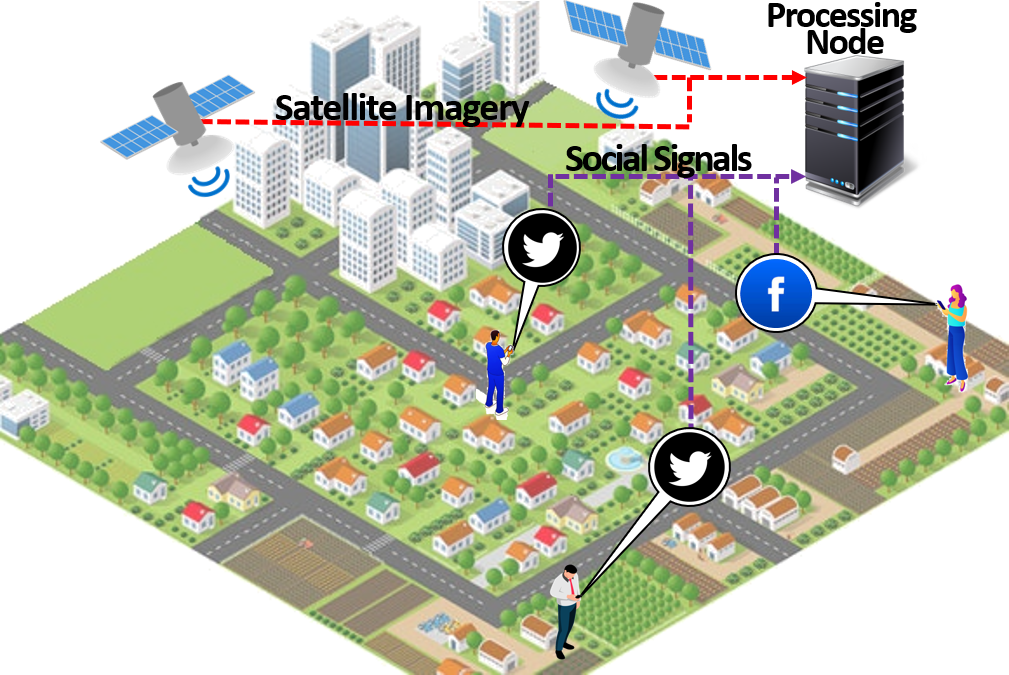}
    \vspace{-0.25in}
    \caption{Scenario of integrated social sensing and satellite-based environmental monitoring}
    \vspace{-0.1in}
    \label{fig:satsocial}
\end{figure}

Several recent studies in SPS have focused on applications integrating satellite-based remote sensing with social media and crowdsensing for capturing a wide range of visual features of the objects residing on the earth's surface. Examples of such applications include urban land usage classification~\citep{chi2017novel}, predicting
the poverty in underdeveloped areas~\citep{zhao2020remote}, post-disaster damage assessment~\citep{huang2017cloud}, risky traffic location identification~\citep{zhang2018risksens}, and flood inundation mapping~\citep{rosser2017rapid}. Figure~\ref{fig:satsocial} exemplifies an integrated social sensing and satellite-based environmental monitoring scheme for analyzing human mobility in urban
areas~\citep{shao2021urban}. Harnessing the mutual efforts of human sensors and physical sensors installed on satellites results in: i) a more pervasive and fine-grained representation of the objects residing on the earth's surface~\citep{zhang2018risksens}, ii) a reduction of their individual weaknesses (e.g., slow update interval of satellites, poor location accuracy of social sensing)~\citep{zhang2016robust}, iii) localized and real-time information for closely monitoring the environment, which is helpful for applications involving emergency response, smart cities, and environmental hazards~\citep{ghamisi2018multisource}, and iv) a greater spatial resolution, which is crucial for applications like land cover classification, distinguishing urban-rural regions, damage assessment, target identification, and geological mapping~\citep{chi2017novel}. 

The fusion of social sensing with empirical measurements from satellite-based remote sensing has opened opportunities for various interesting SPS applications. For example, Zhang \textit{et al.} developed RiskSens, a multi-view learning approach to identify locations with high traffic risk by combining social media data with satellite imagery data~\citep{zhang2018risksens}. Chi \textit{et al.} proposed Crowd4RS, a land usage and land cover classification scheme that combines satellite images in urban areas with geo-tagged social media photos for a more localized and fine-grained analysis~\citep{chi2017novel}. Rosse \textit{et al.} designed a framework to infer flood inundation levels on different terrains by melding geo-tagged images from social media, optical satellite imagery, and high-resolution terrain mapping using a Bayesian statistical model~\citep{rosser2017rapid}. Wang \textit{et al.} presented an early warning system that fuses Twitter data with historical remote sensing data for detecting and predicting weather-driven natural disasters in near real-time~\citep{wang2018fusing}. 
A Twitter-driven remote sensing approach has been developed to convert geo-tagged tweets into high-resolution raster images and integrate them with satellite-based nighttime lights to infer socioeconomic activities~\citep{zhao2020remote}. Another study has presented a framework to incorporate multi-sourced data from social media, remote sensing, and online databases through spatial data mining and text mining for post-disaster damage assessment~\citep{huang2017cloud}. More recently, an integrated crowdsensing and remote sensing scheme has been proposed that combines remote sensing imagery and mobile phone positioning data for urban land usage mapping ~\citep{ghamisi2019multisource}. By exploiting the collective benefits of social sensing and satellite-based environmental monitoring, the above schemes facilitate a fine-grained interpretation of the earth's geological features.

\subsection{Anomaly detection using social airborne sensing (SAS) and social vehicular sensor networks (S-VSN)}
\subsubsection{Social airborne sensing (SAS)}
Social airborne sensing (SAS) is progressing as a new SPS application domain where social signals are used to dispatch unmanned aerial vehicles (UAVs) for perceiving anomalous occurrences in time-sensitive applications (e.g., disaster response, wildfire monitoring)~\citep{rashid2022seis}. Figure~\ref{fig:sas} in Section~\ref{sec:intro} illustrates the concept of representative SAS schemes~\citep{terzi2020towards}. SAS is motivated by the agility and empirical sensing capabilities of UAVs fitted with physical sensors (e.g., camera, LiDAR, thermal scanner)~\citep{casbeer2005forest} and the ubiquity of social data platforms (i.e., social media and crowdsensing). Thus, SAS attempts to leverage the collective benefits of UAVs and social signals to provide a more rapid response and wider sensing scope than other SPS approaches (e.g., approaches that use satellite imagery or fixed sensors like surveillance cameras). Specifically, a more rapid and timely data acquisition can be delivered by SAS, especially in critical scenarios such as search and rescue missions, post-disaster response and recovery, and tracking potential suspects around crime scenes. 

An SAS system collects and analyzes data from social media and crowdsensing platforms to locate probable events of interest (e.g., a person injured on a roadside, an area getting flooded, or buildings damaged by an earthquake)~\citep{terzi2020towards,rashid2019sead}. Afterward, UAVs are selectively dispatched to the extracted locations using various resource management policies (e.g., game theory, supply chain management, and reinforcement learning) to verify the authenticity of the event reports using their onboard physical sensors and augment the knowledge acquisition. Examples of SAS frameworks from recent literature include: i) a path cheapest arc-based SAS scheme that incorporates calls for help from Twitter and dispatches UAVs for search and rescue missions~\citep{terzi2020towards}; ii) a semantic web and machine learning-based SAS design for disaster management in urban areas~\citep{sukmaningsih2020proposing}; iii) a correlation-driven SAS solution for conducting disaster damage assessment in the aftermath of hurricanes~\citep{yuan2018integration}; and iv) a spatiotemporal-aware SAS framework that identifies latent correlations among reported event locations to dispatch UAVs selectively~\citep{rashid2019collabdrone}.  

\subsubsection{Social Vehicular Sensor Network (S-VSN)}
While SAS schemes offer pervasive and accurate information retrieval in critical scenarios, they still require dedicated UAVs, which are expensive and scarce resources having limited flight times~\cite{rashid2021heterosas}. On the other hand, vehicular sensor networks (VSNs) have matured into a dependable networked sensing paradigm for vigilance and situational awareness along roadways that uses cars equipped with physical sensors (e.g. dashboard cameras) to opportunistically identify event occurrences (e.g., accidents on roads)~\citep{zhang2008efficient}. Harnessing existing vehicular infrastructure does not require additional dedicated sensing equipment, which in contrast to UAVs, is more unobtrusive and reduces deployment cost and time since dedicated agents are not required. However, one limitation of traditional VSNs is that the information collected by vehicles is restricted to only those regions traversed by car drivers, restricting the scope of sensing for VSNs and their adaptability in unraveling new events.

\begin{figure}[!htb]
    \centering
   \vspace{-0.05in}
    \includegraphics[width=8cm]{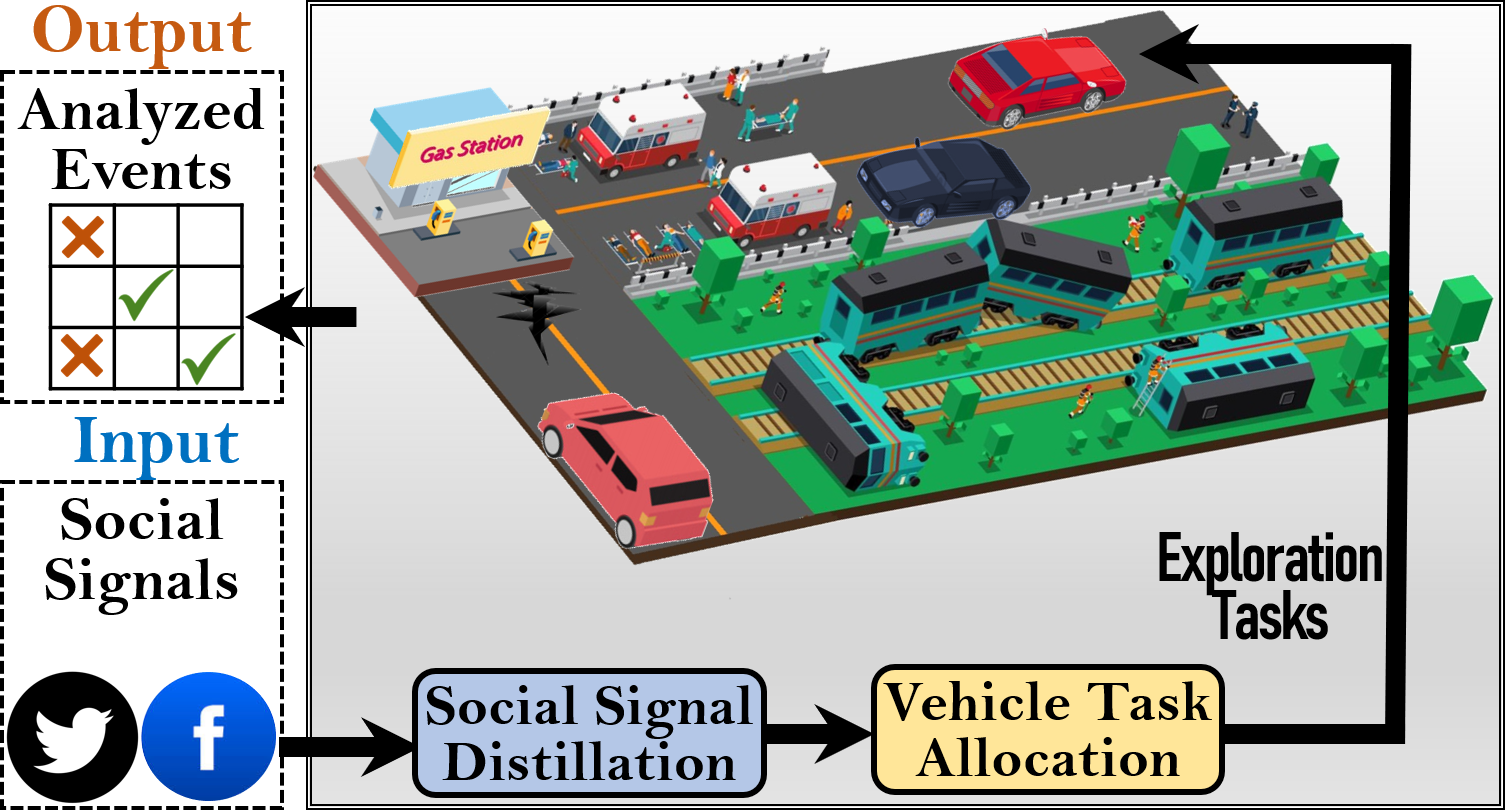}
    \caption{The concept of social vehicular sensor networks (S-VSN)}
    \vspace{-0.1in}
    \label{fig:svsn}
\end{figure}

To this end, an integrated SPS paradigm, namely social vehicular sensor network (S-VSN), has recently been studied to integrate social sensing with existing ground-based VSN to provide more scalable and widespread anomaly detection~\citep{rashid2019socialcar,rashid2020dasc}. Figure~\ref{fig:svsn} shows the concept of an S-VSN scheme where social media users report events of interest~\citep{rettore2019vehicular}. A social signal distillation model analyzes the reports to determine the locations of the events, while a vehicular task allocation model assigns exploration tasks for car drivers to travel to specified locations and analyze the events using car sensors. 

By augmenting the outreach of vehicular sensors with the ubiquity of social sensors, S-VSNs attempt to provide widespread sensing coverage and greater sensing accuracy than standalone VSNs. In specific scenarios, such as identifying risky traffic regions or discovering essential resources in the aftermath of a disaster in large areas (e.g., locating gas availability at gas stations), an S-VSN might be more feasible than an SAS. Recent examples of S-VSN frameworks include: i) a community-aware S-VSN architecture for road traffic anomaly detection~\citep{qiu2018community}; ii) an S-VSN system for performing accident investigation in smart cities~\citep{rettore2019vehicular}; and iii) a road damage-aware S-VSN scheme that uses a Markov Decision Process (MDP)-based damage discovery scheme to locate roads affected by damage after a disaster~\citep{rashid2020dasc}. 

\subsection{Automatic license plate recognition using crowdsensing and physical sensors}
\begin{figure}[!htb]
    \centering
    \includegraphics[width=8.5cm]{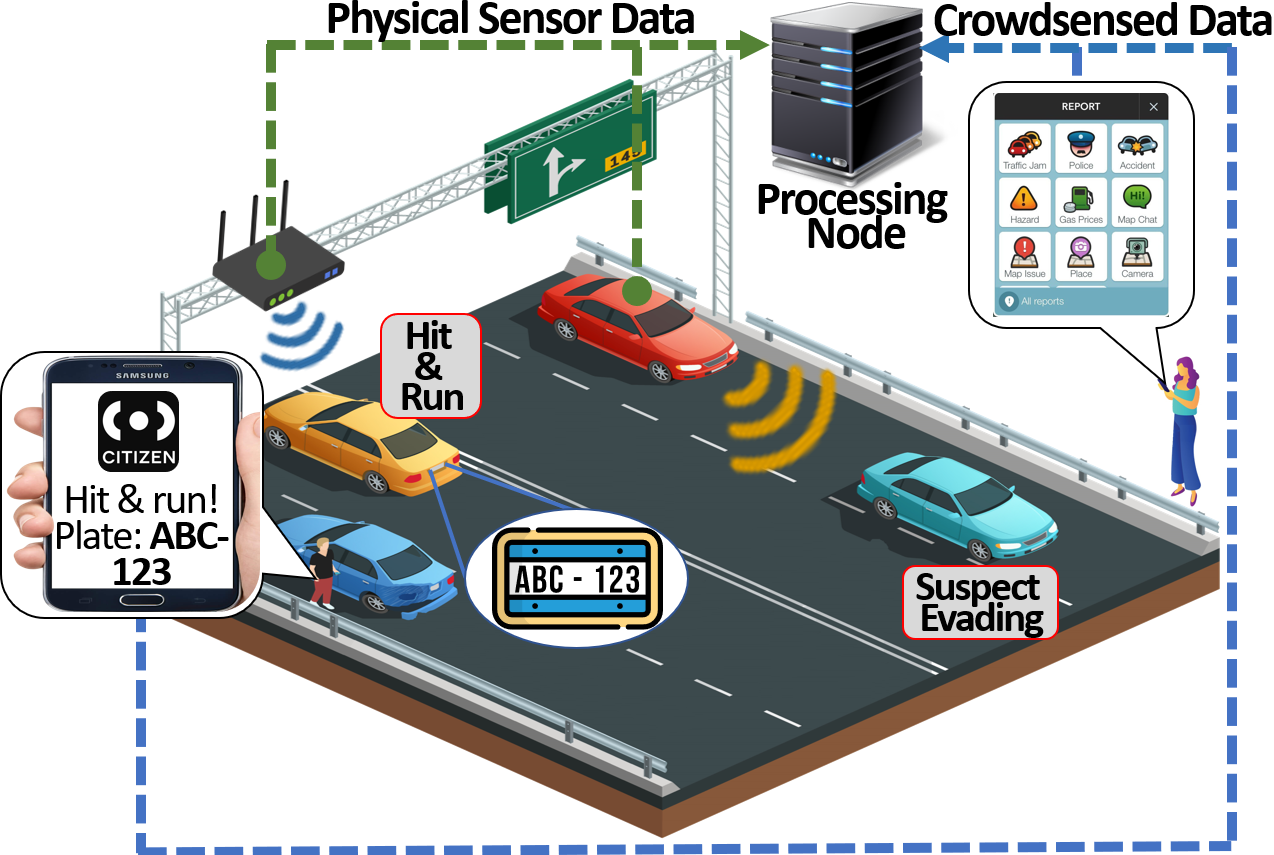}
    \caption{Overview of automatic license plate recognition using crowdsensing and physical sensors}
    \label{fig:alpr}
\end{figure}

One recent SPS application domain is automatic license plate recognition (ALPR) based on crowdsensing (e.g., smartphone apps) and physical sensors (e.g., roadside units, vehicular sensors, and smartphone sensors). Figure~\ref{fig:alpr} illustrates an SPS-based ALPR application where information from traffic monitoring devices (e.g., roadside cameras and dashboard cameras) are melded with human inputs from crowdsensing apps (e.g., Citizen, Waze, Neighbors) to track down the plate number of a potential suspect's vehicle evading from a crime scene (e.g., a hit-and-run)~\citep{ang2018deployment}. The analytics are typically conducted using deep-learning algorithms~\citep{zhangedgebatch, ang2018deployment}. Thus, observations contributed by drivers, passengers, and commuters on roads might be integrated with knowledge from hardware sensors to narrow down searches by law enforcement personnel and swiftly locate the whereabouts of perpetrators.

One crucial concern of SPS-based ALPR applications is their real-time requirements, where plate detection tasks are expected to be accomplished within certain time bounds in resource-constrained environments (e.g., the devices might have limited network bandwidth). Existing standalone ALPR approaches primarily focus on analyzing large volumes of video footage data collected from surveillance cameras and stored in the cloud platforms~\citep{zhang2017enhancing}. However, such schemes often introduce a non-trivial amount of data transmission delay to offload the videos to the cloud, which is not favorable for the real-time car plate detection application. More recently, there is a growing development of ALPR schemes that harness crowdsensing combined with existing vehicular sensors and IoT devices (e.g., vehicles equipped with dash cameras and smart devices owned by citizens) to form a city-wide video surveillance network that tracks moving vehicles using the automatic license plate recognition (ALPR) technique~\citep{du2012automatic}. Zhang \textit{et al.} developed EdgeBatch, an SPS-based ALPR task management framework where reports about license plates of probable suspects from concerned citizens in crowdsensing apps are combined with inputs from IoT sensors (e.g. surveillance cameras) using collaborative edge computing resources to detect the license plates~\citep{zhang2019edgebatch}. Trottier \textit{et al.} presented the concept of a dashboard camera and crowdsensing platform-driven ALPR scheme for smart cities where video footage from dashboard cameras is analyzed by image processing algorithms and further augmented with inputs from crowdsensing participants through an app to recognize the number plates~\citep{trottier2014crowdsourcing}. 

Despite their usefulness, ALPR approaches also instill privacy concerns in the collaborative sensing context of SPS applications. For example, car drivers might not be willing to share the metadata from their devices to the cloud for fear that such data may reveal their private information (e.g., location, speed, and driving behavior). With concerns about user privacy, Alcaide \textit{et al.} proposed a privacy-aware ALPR scheme that maintains confidentiality of the users' data and prevents unauthorized usage of private devices that are used for capturing and recognizing images of plate numbers~\citep{alcaide2014privacy}. A privacy-aware ALPR scheme has been proposed that masks and protects the identity of the owners of license plates recognized using data from crowdsensing apps and roadside monitors~\citep{yan2011crowdpark}. By exploiting the knowledge from crowdsensing and physical sensors, SPS-based ALPR applications aid in tracking down potential criminals on roads~\citep{zhangedgebatch}.

\subsection{Situational awareness using social media and crowdsensing melded with IoT (Social/CrowdIoT)} \label{sec:crowdiot}
\begin{figure}[!htb]
    \centering
    \includegraphics[width=8cm]{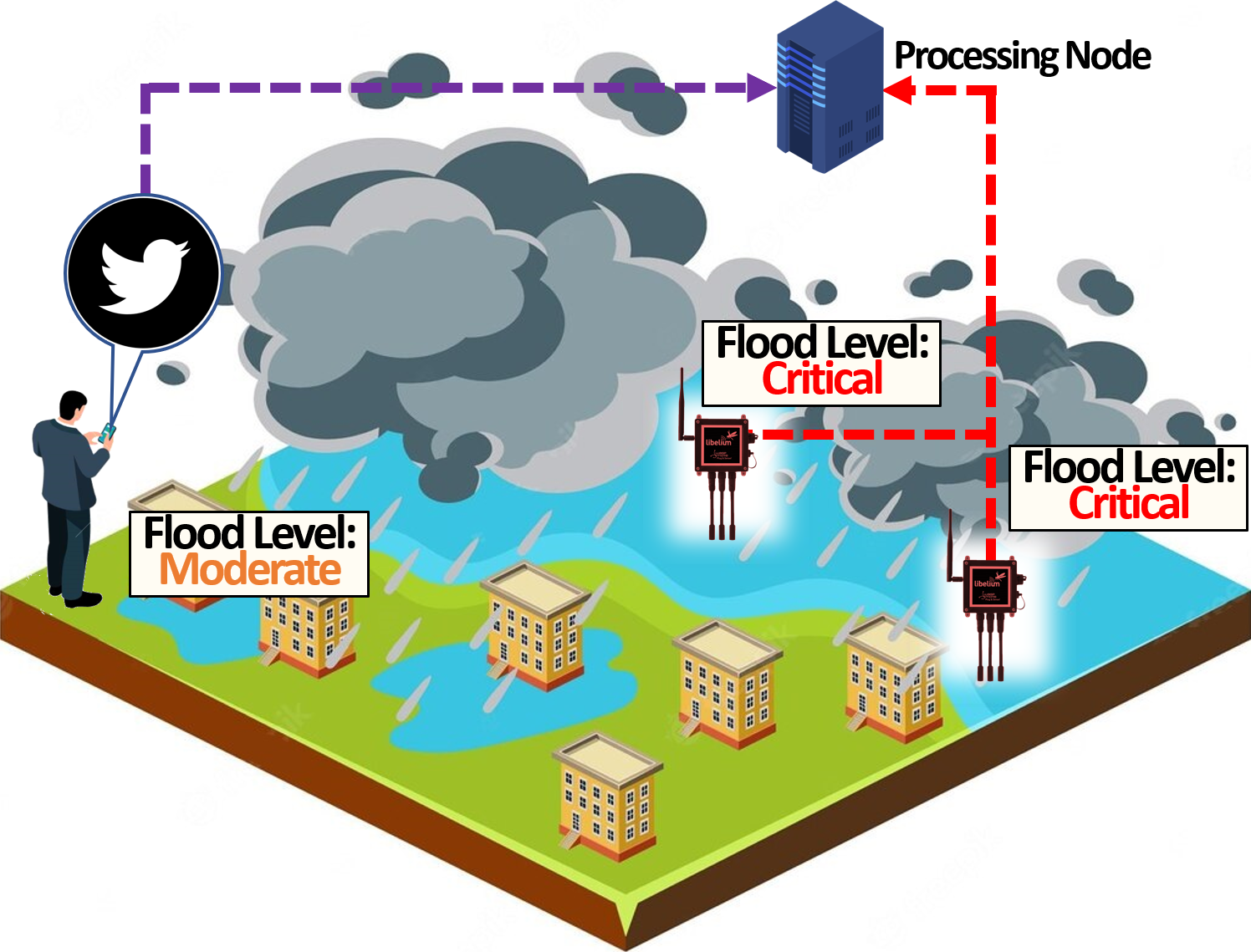}
    \caption{Overview of situational awareness with SocialIoT}
    \label{fig:soccrowdiot}
\end{figure}

The prevalence of IoT alongside social media and crowdsensing has opened new domains for situational awareness in SPS. Examples of such applications include real-time crowd density measurement, search and rescue operations, and urban anomaly detection~\citep{kucuk2019crowd, atzori2010internet,zanella2014internet, brabham2013crowdsourcing}. By integrating social media and crowdsensing with the IoT paradigm, the emerging areas of SocialIoT and CrowdIoT, respectively, can achieve results beyond what is possible with traditional standalone situational awareness approaches. Figure~\ref{fig:soccrowdiot} illustrates a SocialIoT-based situational awareness application where information from Twitter and IoT-enabled flood measurement sensors can be combined to estimate the density of flood~\citep{mirza2022improving}. The following subsections discuss a few variants of SocialIoT and CrowdIoT.

\subsubsection{Integrated Social Media Sensing and IoT (SocialIoT)-based Indoor Localization and Tracking}
In recent times, there has been a surge in SPS applications that focus on indoor localization based on contextual information provided on social media and raw signals from IoT devices. While GPS provides fairly accurate outdoor location tracking, the applicability of GPS for indoor tracking is limited primarily due to the inaccessibility of satellite signals inside confined spaces and lower degrees of precision. As such, accurate indoor localization schemes require additional infrastructure support (e.g., ranging devices) or extensive training before system deployment (e.g., WiFi signal fingerprinting). In indoor localization, networks of IoT devices are used to track people or objects in confined places where GPS and other satellite technologies usually lack precision or fail entirely, such as inside multistory buildings, airports, alleys, parking garages, and underground locations. Location-based services, such as targeted advertisement, geosocial networking, and emergency services, are becoming increasingly popular for mobile SPS applications~\citep{jun2013social,liu2010indoor}.

In order to help existing localization systems to overcome their limitations or enhance their accuracy, approaches have been developed that combine social media sensing with IoT for accurate location tracking indoors. For example, a scheme called Social-Loc has been proposed that integrates the physical traces of an individual posted through social media (e.g., check-ins to a particular shop in a shopping mall) with RSSI signals from WiFI routers to potentially derive the exact location of individual users within a building~\citep{jun2013social}. Chu \textit{et al.} designed SBOT, a social media and sensor network-driven indoor localization scheme which combines user statuses and updates posted through social media (using text mining techniques) with telemetry data from smartphone sensors (e.g., altitude, speed, and heading of the users) to pinpoint the location of the users~\citep{chu2020sbot}. Liu \emph{et al.} proposed a social-driven IoT system consisting of backpacks equipped with 2D laser scanners and inertial measurement units augmented with historical social network traces of the users to perform indoor localization and visualization of complex environments such as staircases or corridors~\citep{liu2010indoor}. However, with the increasing facilities for geo-locating people using their digital footprints, concerns for individuals' privacy also prevail. As we will discuss later in Section V-E, the metadata obtained from the social and physical sensors in SPS for locating people exposes the risks of revealing their private information. A few privacy-preserving SocialIoT schemes have been developed which aim to protect people's identities while geo-locating them indoors~\citep{hamza2020privacy,perera2015big}.


\subsubsection{CrowdIoT-based Context Awareness}
Several exciting CrowdIoT applications have emerged that are crucial to society's well-being, including criminal identification and disaster response~\citep{dunphy2015crowdsourcing}. Dunphy \textit{et al.} proposed an integrated crowdsensing and CCTV-based video surveillance framework where surveillance footage collected from CCTVs spread across a city is assigned to Amazon MTurk participants to tag instances of abnormal occurrences in real-time (e.g., traffic accidents, crimes)~\citep{dunphy2015crowdsourcing}. 
Abu \emph{et al.} designed an integrated risk assessment framework using crowdsensing apps and fixed urban IoT sensors (e.g., proximity sensors, acoustic sensors, radars,  air quality monitors, etc.) that predicts the possibility of crisis such as  multi-vehicle accidents, major weather events, and large fires\citep{abu2016enhancing}. Vital information from the framework might assist emergency personnel such as firefighters and first responders. Beyond surveillance-centric context awareness applications, stationary CrowdIoT-based SPS schemes are also used for locating regions of adverse weather and climatic conditions. For example, Horita \textit{et al.} developed a flood inundation mapping (FIM) system that integrates crowdsourcing data with data from \textit{in situ} weather radars to infer probable locations of a flood~\citep{horita2018determining}. Thus, building upon the tight integration of crowdsensing and fixed-sensor IoT devices, stationary CrowdIoT solutions (like the ones discussed above) facilitate providing rich context-aware SPS applications.



Another emerging context awareness sub-domain within SPS involves integrating crowdsensing with mobile devices and portable IoT devices, otherwise known as mobile crowdsensing (MCS). Applications integrating mobile sensors with crowdsensing in MCS utilize users with mobile devices capable of data capturing, computation, and communication to collectively share data and extract information to measure, assess, estimate, or predict processes of shared interest \citep{ganti2011mobile, ma2014opportunities,guo2015mobile,wang2014surrogate}. Such mobile devices include smartphones, wearables, and tablet computers equipped with hardware sensors (e.g., GPS, microphones, heart rate monitors) and sufficiently robust processing units (e.g., CPU, FPGA, GPU). The ubiquity of such \textit{``all-in-one"} data acquisition, computation, and communication devices has motivated a good amount of work in developing a wide range of SPS-based urban sensing tools~\citep{zappatore2016using,yan2017cloud,li2018toward}. A few essential applications fueled by mobile crowdsensing include: i) real-time urban crisis reporting where inputs from concerned citizens through smartphone apps and signals from IoT sensors (e.g., proximity sensors) are correlated to located urban crisis~\citep{konomi2015crowd}; ii) risky traffic zone identification where crowdsensed traffic data from dedicated websites are combined with roadside sensor units to locate traffic risks~\citep{li2019enhancing}; iii) gas leakage detection in urban areas in which gas sensors are used to measure unusual gas concentrations and further integrated with knowledge from citizens acquired through crowdsensing apps to identify gas leakage~\citep{akter2020location}; and iv) simultaneous localization and mapping for rescue missions in which reports of potential survivors from smartphone apps are augmented with received signal strength indicator (RSSI) values from WiFi routers
to locate potential survivors in the aftermath of disasters~\citep{kucuk2019crowd}.

In addition to the above critical mobile crowdsensing schemes, there have been significant works on utilizing smartphones and wearable sensors (e.g., sociometric badges, smart glass, fitness trackers, and smartwatches) for less critical applications such as: i) monitoring environmental conditions like noise \citep{zappatore2016using} and air quality \citep{vahdat2018architecture}; ii) assessing 
infrastructural conditions such as traffic congestion \citep{yan2017cloud} and road damage \citep{li2018toward}; and iii) determining most fuel-efficient travel routes~\citep{ganti2010greengps}. The integration of crowdsensing and mobile sensors has also opened up new possibilities for exciting applications in disaster response. Han \emph{et al.} \citep{han2019harnessing} proposed a crowdsensing and mobile-IoT integration model that aims to improve disaster response by using important metrics such as weather conditions, damage reports, and infrastructure accessibility derived from crowdsensing apps and portable devices equipped with RFID technology. Driven by the unification of crowdsensing with sensors contained in mobile devices, mobile crowdsensing schemes aim to provide a more holistic representation of the environment in SPS applications. 

The following section discusses key research challenges prevalent in current SPS applications.
	\section{Fundamental Challenges in SPS} \label{sec:challenges}

\begin{table*}[hbt!]
  \centering
  \caption{Summary of Schemes Targeting the Challenges in SPS and Open Research Questions}
  \scalebox{1}{
	\begin{tabular}{p{1.8cm} p{2cm} p{4.5cm} p{7cm}}
	\toprule
	Challenge & Description & Schemes Targeting Challenge & Open Research Questions\\
	\cmidrule(l){1-4}
	Data Collection Challenge &Locating raw sensor data from numerous social and physical sensors& \citep{nur2015combination,wang2019age,zhang2018light,jagannatha2016structured,zhang2020transres,lai2018deep,wang2014biggerpicture,zhang2020towards,zhang2011sedic,heydon2012bluetooth,johnsen2018application,hull2003bandwidth}& 
	\begin{itemize}
    \item How to systematically locate useful data from inherently \textit{noisy} social and physical signals?
    \item How to gain access to sensing data from privately-owned devices?
    \end{itemize}
	 \\
	\cmidrule(l){1-4}
	Human-Cyber-Physical Interactions Challenge&Handling the complex interactions between the human, cyber, and physical domains &\citep{lee2019optimal,zhang2019heteroedge,sathiyanarayanan2019understanding,rashid2019socialcar,rashid2020dasc}&  
	\begin{itemize}
    \item How to develop a closed-loop system that seamlessly integrates social and physical sensors?
    \item How to explicitly model the roles of human participants as actuators?
    \item How to use physical sensors to validate knowledge contributed by human sensors? 
    \item How to leverage social signals to effectively control physical sensors' performance? 
    \end{itemize}\\
	\cmidrule(l){1-4}
	Device and Data Heterogeneity Challenge&Managing the diversity of the devices and data associated with the social and physical sensors &\citep{shao2018dynamic,gigan2007sensor,scheepers2014virtualization,jun2019dynamic,kirkpatrick2013software,khan2015wireless,oza2005online,gan2001comparison,zhang2018risksens,zhang2019riskcast}& 
	\begin{itemize}
    \item How to apply global policies and control privately owned devices from a central authority perspective? 
    \item How to explicitly consider the heterogeneity of tasks and architectures for devices? 
    \item How to manage the complex interdependence of tasks distributed across multiple devices? 
    \item How to analyze the different types of data that vary across dimensionality? 
    \item How to handle the different rates of data generated by social and physical sensors?
    \end{itemize}\\
	\cmidrule(l){1-4}
	Dependency and Correlation Challenge &Characterizing the dependencies between sources and correlating the collected data &\citep{dey2018fake,asim2019trust,ahn2011new,giridhar2016clarisense+,tsapeli2017non,rashid2019sead}&
	\begin{itemize}
    \item How to model source dependency and data provenance, given the diverse source dependency nature of social and physical sensing? 
    \item How to identify and incorporate implicit correlations within events obtained from social and physical sensors? 
    \item How to explore strong causal relationships between physical and social sensor data?
    \end{itemize}\\
    \toprule
    \end{tabular}
    }
\label{tab:Challengesummary}
\vspace{-0.1in}
\end{table*}
    
    \addtocounter{table}{-1}
    
    \begin{table*}[hbt!]
  \centering
  \caption{Summary of Schemes Targeting the Challenges in SPS and Open Research Questions (Continued)}
  \scalebox{1}{
	\begin{tabular}{p{1.8cm} p{2cm} p{4.5cm} p{7cm}}
	\toprule
	Challenge & Description & Schemes Targeting Challenge & Open Research Questions\\
	\cmidrule(l){1-4}
	Privacy Challenge &Mitigating privacy issues arising from the integration of social and physical sensors &\citep{toch2012personalization,liu2019privacy,vance2018privacy,ganti2008poolview,li2009privacy,al2015internet,toch2012personalization,li2009privacy}& 
	\begin{itemize}
    \item How to develop robust privacy-conserving schemes to prevent the malicious exploitation of complementary information from social and physical sensors? 
    \item How to design integrated privacy-aware SPS platforms to concurrently consider the data heterogeneity and protect sensitive user information?
    \end{itemize}\\
	\cmidrule(l){1-4}
	Dynamics Challenge &Adapting to the interrelated dynamics from the social and physical realms&\citep{rashid2020compdrone,zhang2017towards,rashid2019socialcar,li2019introduction,wang2013exploitation}&  
	\begin{itemize}
    \item How to handle interrelated dynamics induced by the fusion of social and physical domains?
    \item How to adapt to the dynamics from the social domain which impacts the performance of physical sensing? 
    \item How to adapt to the physical world dynamics which affect the performance of both the social and physical sensors?
    \end{itemize}\\
\toprule
    \end{tabular}
    }
\label{tab:Challengesummary2}
\vspace{-0.1in}
\end{table*}

This section highlights a few fundamental open challenges in the interaction and integration between social and physical sensing in SPS. Table~\ref{tab:Challengesummary} presents a comprehensive summary of the challenges. In particular, the first column of the table indicates the challenge, which ranges across data collection, human-cyber-physical interactions, device and data heterogeneity, dependency and correlation, privacy, and dynamics. The second column provides a brief description of the challenge. The third column provides references to schemes from current literature targeting the challenge. Lastly, the fourth column presents a set of possible open research questions to solve to address the challenge. In the following subsections, we discuss each challenge in detail and highlight the measures to address the challenges in current literature and their shortcomings.

\subsection{Data Collection Challenge}
Before valuable knowledge can be interpreted in SPS, the relevant data must first be located, extracted, and organized. Thus, one of the critical challenges in SPS lies in simultaneously harvesting the raw sensor data from myriads of social and physical sensors~\citep{stieglitz2018social,wang2011optimizing,zhang2019sparse}. 

\begin{figure}[!htb]
    \centering
   \vspace{-0.05in}
    \includegraphics[width=8.5cm]{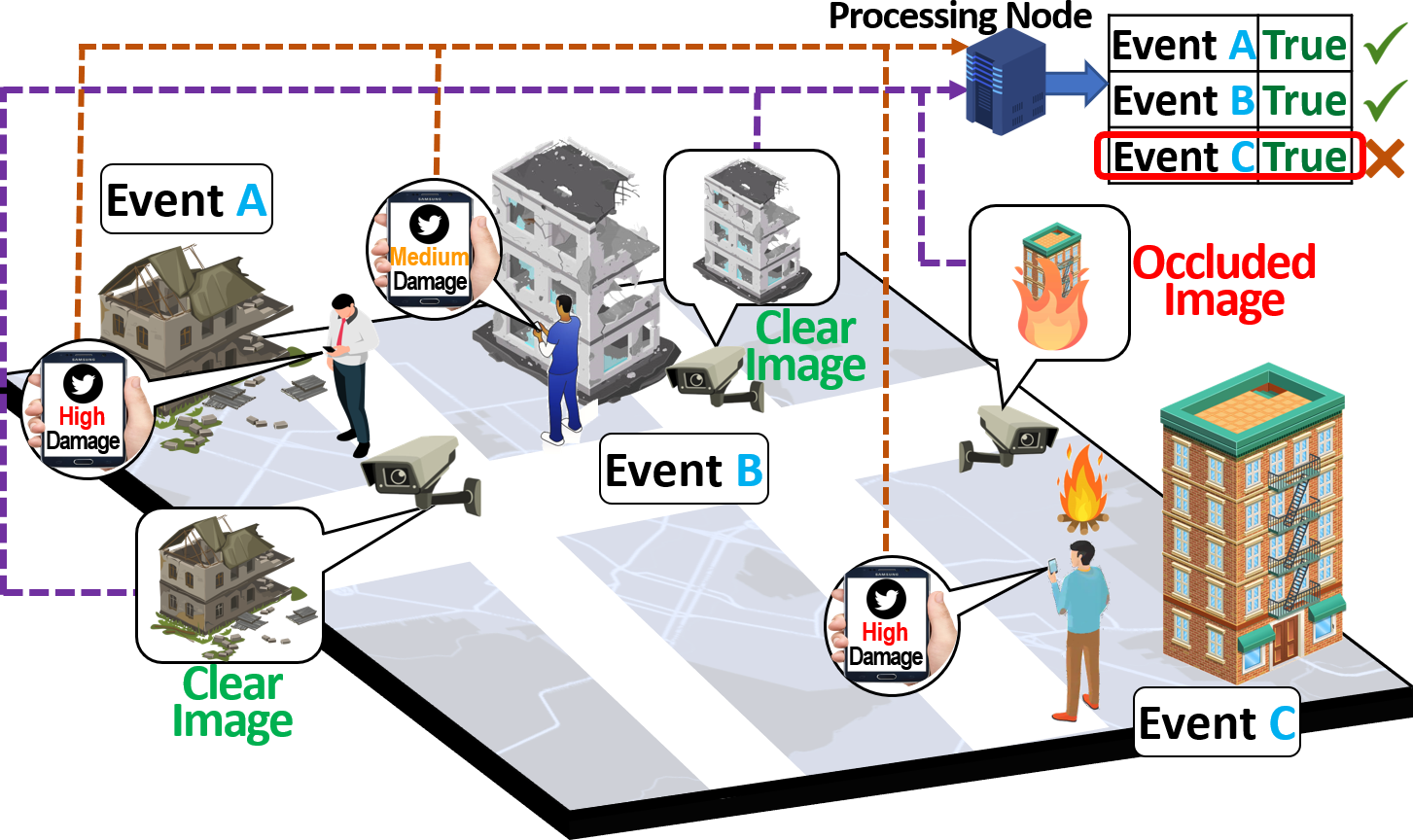}
    \vspace{-0.05in}
    \caption{Example of data collection challenge in SPS Applications}
    \label{fig:datacollect}
\end{figure}

The first obstacle in data collection is to systematically locate useful data from the inherently \textit{noisy} social and physical signals. In knowledge discovery from social data platforms (e.g., social media websites), traditional search techniques use keywords to query for the related data~\citep{stieglitz2018social}. However, such searches might return a considerable amount of reports of unrelated incidents (i.e., noisy data) alongside the relevant ones. On the other hand, hardware sensors are susceptible to several types of characteristic noise that cause deviation in the data capture (e.g., satellites images might have low resolutions, and drifts in GPS data might record incorrect location information)~\citep{tsai2014framework,sundvall2006fault}. When combined in an SPS setting, the noises originating from the social and physical sensors can develop a degree of interdependence among each other, causing difficulty in collecting useful data. For example, let us consider an integrated social media and surveillance camera-based damage assessment application~\citep{bartoli2015novel}. Unreliable human sensors often incorrectly report sites of damage. If surveillance cameras capture images with incorrect perspectives (e.g., due to occlusions or faraway positions), they might not reveal the true state of the damage, and the reliability of the sources might not be validated correctly (e.g., reliable sources might get flagged as malicious). Figure~\ref{fig:datacollect} shows an example of such an application where events A and B are \textit{true} reports of damaged sites, but event C is a \textit{false} report (i.e., a person posts disinformation indicating that the building is on fire which in reality is not). Due to being occluded by a set of burning logs and positioned far away from the building, the surveillance camera at event C might capture a perspective that can cause a computer vision (CV) algorithm to `think' that the building is actually on fire, resulting in the sensing framework to consider the unreliable source to be trustworthy.

Existing literature on social sensing has proposed methods to overcome the noise from social data platforms with techniques such as machine learning (ML)~\citep{nur2015combination}, artificial neural networks (ANNs)~\citep{jagannatha2016structured}, 
estimation theory~\citep{wang2019age}, and adaptive sampling~\citep{zhang2018light}.
Studies on physical sensors have proposed methods to reduce sensor noise using approaches like image enhancement with super-resolution~\citep{zhang2020transres}, deep learning-driven noise reduction~\citep{lai2018deep}, and graph neural network-based data extrapolation~\citep{wang2014biggerpicture}. However, such standalone approaches fail to address the intrinsic interdependence between the noise from social and physical signals in SPS, which is non-trivial to quantify and model.

The second obstacle is gaining access to sensing data from devices owned by individuals. While there is an abundance of connected devices that are able to perform a wide range of data capture, computation, and communication tasks, a significant number of them are privately owned (e.g., smartphones, IoT devices, surveillance cameras)~\citep{johnsen2018application}. Consequently, gaining access to such sensors' data is difficult primarily because the individual entities might not be willing to share their personal devices due to reasons such as inconvenience, draining of battery on mobile devices, usage of cellular data, and privacy concerns~\citep{wang2009survey}. 

Recent literature has presented several solutions like: i) privacy-aware schemes such as game-theoretic task allocation~\citep{zhang2020towards} and non-invasive distributed private data collection~\citep{zhang2011sedic}; ii) energy-preserving data transmission schemes such as Bluetooth low energy~\citep{heydon2012bluetooth}; and iii) bandwidth-conserving data sharing tools such as signal compression~\citep{johnsen2018application} and hop-by-hop flow control~\citep{hull2003bandwidth}. These approaches might potentially help to convince people to provide access to their devices for obtaining sensor data. However, beyond the willingness of people to share their personal devices, the devices in SPS might be unavailable for capturing or providing access to the data. For example, a user might be using her smartphone to play video games or watch videos, making the device unavailable for capturing images and processing them efficiently. Therefore, collecting data from the social and physical realms that direct to the appropriate information remains an open challenge in SPS.

\subsection{Human-Cyber-Physical Interactions Challenge}
In SPS, one elemental challenge is handling the complex interactions between the human, cyber, and physical (HCP) components when integrating social sensing with physical sensing. As events in the real world play out, human and physical sensors are expected to spontaneously contribute knowledge through the social and physical data platforms to recover the truthful states of real-world occurrences. Given this basis, developing a closed-loop system that seamlessly integrates the social and physical sensing paradigm is crucial. 

In such a closed-loop system, the social and physical sensors effectively communicate and complement each other to accomplish the assigned sensing tasks jointly. Existing research on human-computer-interactions (HCI) has explored the need for designing effective interfaces to connect the human and cyber worlds, which include examples such as web interfaces, mobile applications, online forms and survey questionnaires, virtual reality (VR), and motion capture~\citep{wich2015enhanced}. In recent times, there has been a surge in research on cyber-physical systems (CPS), which explores the interactions between the cyber and physical worlds with a focus on the problems in sensing, computation, and control of a CPS system~\citep{zeng2020survey}. Recent studies in CPS have proposed techniques such as embedding human intelligence into cyberspace and augmented reality-driven assistive technology for humans to reduce the gap between the human and cyber worlds~\citep{hu2012review,zhang2020crowd}.
However,  handling the HCP interactions in SPS is much more complex and challenging than the problems studied by existing HCI and CPS research.

\begin{figure}[!ht]
    \centering
    \includegraphics[width=8.75cm]{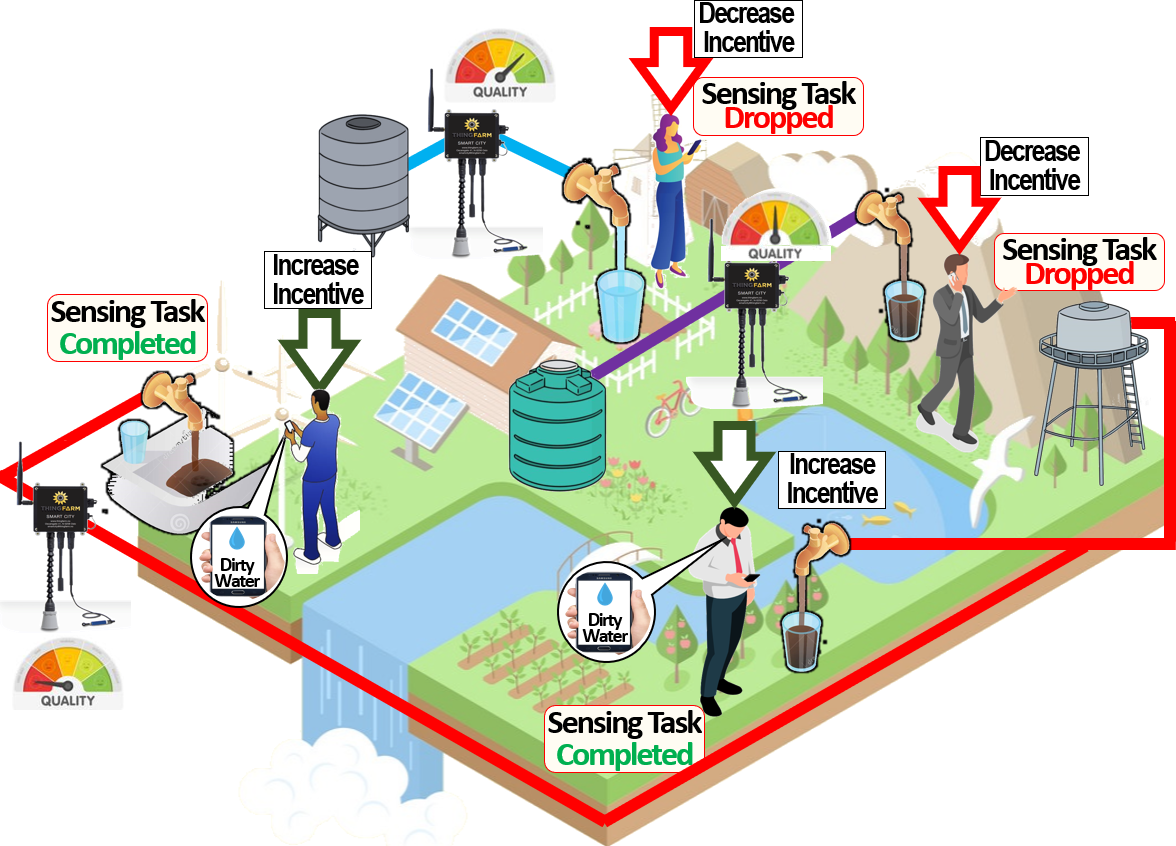}
    \vspace{-0.05in}
    \caption{Example of human's role as actuators in SPS}
    \vspace{-0.05in}
    \label{fig:hcp}
\end{figure}

While human users typically act as sensors in SPS applications, they must also carefully consider their roles as actuators. Let us consider an example in Figure~\ref{fig:hcp}, which shows a smart water monitoring application where crowdsensed water quality measurement is combined with physical water quality sensor data~\citep{fascista2022toward}. Here, crowdsensing participants act as actuators. If the participants do not contribute data of sufficient quality (i.e., not enough reliable data or low participation level), incentives can be applied to encourage them to provide better-quality data. The incentives serve as control signals, and the participants act as actuators. Upon receiving higher incentives, the participants might potentially take a response/action in the physical world by: i) collaborating to contribute more data; ii) validating the data of their peers; iii) or encouraging more people to participate by referring them to use the app~\citep{peng2015pay}. Thus, the incentive serves as a signal from the cyber world (i.e., through smartphone apps) to control response in the human world (i.e., the human participants). When humans receive the incentives, they respond in the physical world (i.e., collect and contribute higher quality data). Such an adaptive closed-loop system requires a careful design that systematically models the complex HCP interactions.

Current literature has proposed methods to develop closed-loop systems encompassing various sensors (e.g., cameras, GPS sensors), actuators (e.g., robotic arms, motorized doors), and controllers (e.g., proportional–integral–derivative (PID) controllers, fuzzy logic controllers, reinforcement learning) for establishing effective cooperation between them using techniques such as linear quadratic Gaussian (LQG) control~\citep{lee2019optimal}, supply chain theory~\citep{zhang2019heteroedge}, and blockchain-based smart contracts~\citep{sathiyanarayanan2019understanding}. However, the closed-loop challenge at the intersection of human, cyber, and physical spaces in SPS has not been fully addressed by existing research for several reasons. First, current solutions often do not explicitly model the human participants as actuators, which is a crucial feature of SPS applications. Second, current literature on incentive design in crowdsensing frameworks has not addressed how to use the physical sensors to validate the information contributed by human sensors. Third, existing approaches have not fully explored measures to leverage social signals to effectively control the performance of the physical sensors. Last, current solutions have not explicitly considered the joint dynamic nature of the human, cyber, and physical worlds to tightly coordinate their interactions. As such, addressing the HCP interaction prevalent in SPS systems remains an open challenge.

\subsection{Device and Data Heterogeneity Challenge}
\begin{figure}[!htb]
    \centering
    \includegraphics[width=8cm]{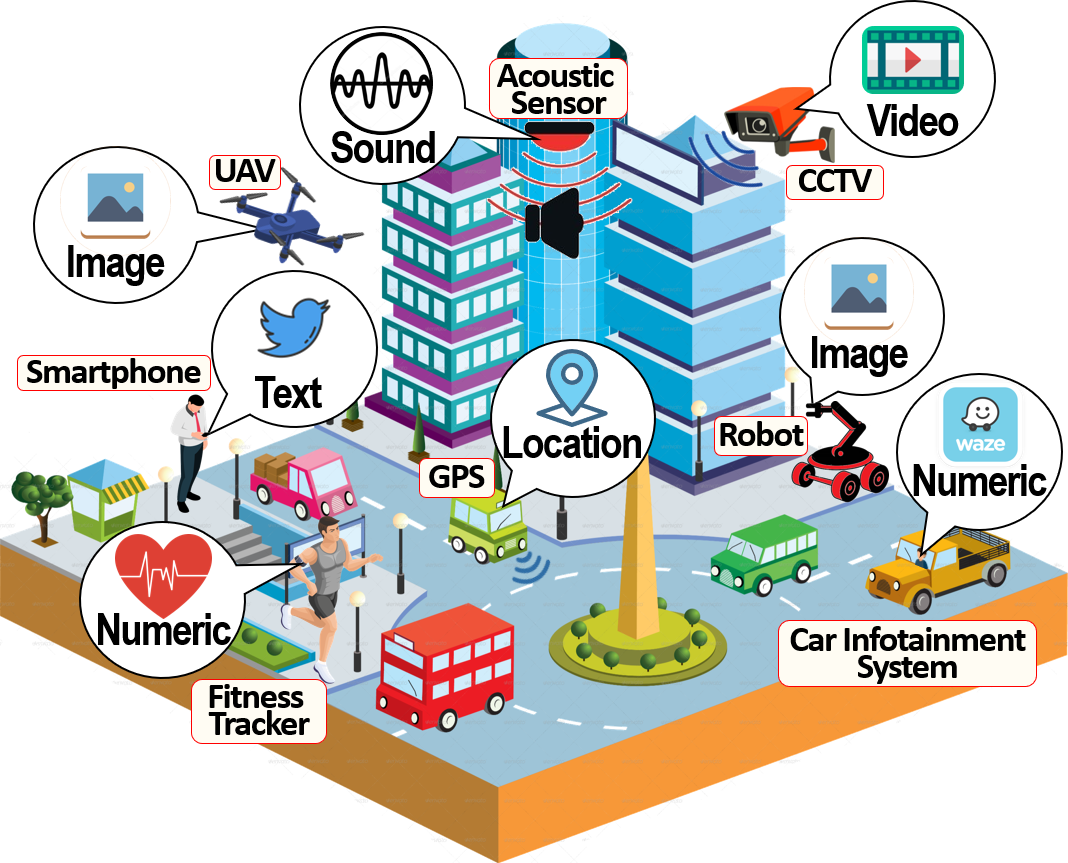}
    \caption{Scenario of data and device heterogeneity challenge in SPS Applications}
    \label{fig:hetero}
\end{figure}

While the abundance of physical and social sensors in SPS provides a rich influx of knowledge across various sensing applications, an inherent challenge in SPS lies in managing the diverse range of devices involved in the sensing process and the different types of data they generate. We deem this challenge as \textit{device and data heterogeneity}. Figure~\ref{fig:hetero} demonstrates a scenario of the data and device heterogeneity challenge in the context of a smart city~\citep{guo2017csf}. We can observe that multiple users and devices generate data in various formats such as text, images, sound, video, numeric, and geo-location. Such multi-modal data is non-trivial to analyze and interpret as we shall see.

As identified in Section~\ref{sec:enabling}, SPS applications are centered around a diverse collection of devices that encompasses data acquisition, communication, and computation. In particular, the physical sensing components rely on the capabilities of hardware sensing devices (e.g., cameras, UAVs, and robots), while the social sensing components obtain observations from human sensors through crowdsensing and social media by implicitly leveraging user devices (e.g., connected tablets, laptops, and smartphones). Such devices have distinct characteristics in terms of sensing and computation capabilities, sensitivity, power requirements, frequency of data capture, communication protocols, access control and authentication methods, and runtime environments~\citep{zhang2019heteroedge,chu2016data,zhang2019transland,zhang2020transres,shang2019towards}, which often presents a unique difficulty in managing them in SPS applications. For example, in the SAS application of Figure~\ref{fig:sas} in Section \ref{sec:intro}~\citep{rashid2019collabdrone}, smartphones capture human observations and send them to social media platforms which are then used to dispatch UAVs to recover the veracity of the reports. Standalone social or physical sensing applications are unlikely to have such diverse devices working together. As such, little work has been done in earlier research to bridge the knowledge gap in SPS and construct a unified framework that can efficiently manage such diverse devices.

A few efforts have attempted to mitigate device heterogeneity in sensor networks and distributed systems primarily using abstraction-based approaches such as: i) sensor emulation, device clustering~\citep{shao2018dynamic}, and sensor abstraction layer~\citep{gigan2007sensor} for data acquisition devices; ii) containerization~\citep{scheepers2014virtualization} and dynamic binary translation~\citep{jun2019dynamic} for computation devices; and iii) software-defined networking~\citep{kirkpatrick2013software} and sensor network virtualization~\citep{khan2015wireless} for communication devices. However, in the context of SPS, existing solutions are inadequate in addressing the device heterogeneity challenge due to several reasons: i) the devices in SPS are mostly privately owned (i.e., smartphones, IoT devices), which makes it hard for an SPS application to apply global policies and control the devices from a central authority perspective \citep{zhang2019heteroedge} (e.g., it might not be possible to install a middleware application on a personal device); ii) the extent of heterogeneity of the devices in SPS is more evident due to the added heterogeneity of tasks and architectures which current solutions overlook~\citep{zhang2019heteroedge}; and iii) the devices in SPS often have complex interdependence of the tasks~\citep{zhu2019delay}, which existing solutions might not preserve~\citep{wei2019computational}.  

Beyond the diversity of the devices, the social and physical sensors in SPS typically generate data that widely vary across modalities and formats. For example, the input data type can range across text, image, location, audio, and video~\citep{birke2014big}, and each type can further encompass different dimensionality, making the data heterogeneity even more pronounced~\citep{zhai2014emerging}. For example, for image data, the dimensionality can be edges, corners, blobs, and ridges, while for text data, the dimensionality can be document frequency and \textit{n}-grams~\citep{khanina2012scale}.  
Existing methods for mitigating data heterogeneity include data fusion schemes such as bagging and boosting~\citep{oza2005online}, deep learning (DL)-driven data fusion~\citep{gao2020survey}, covariance intersection~\citep{gan2001comparison} as well as other statistical and machine learning methods such as dimensionality reduction~\citep{renard2009dimensionality}, multi-view learning~\citep{zhang2018risksens,zhang2019riskcast}, and feature concatenation~\citep{gao2020survey}. Despite their effectiveness in standalone sensing applications, current approaches fail to address the data heterogeneity issue in SPS due to the inherent complexity injected by the different data rates generated by the social and physical sensors in SPS applications. The diverse sensors in SPS are known to produce data at different frequencies, rendering existing solutions infeasible~\citep{misra2020iot,mourtzis2016industrial}. Consequently, versatile data management schemes must be developed to withstand the heterogeneity of data in SPS and interpret knowledge from the social and physical signals. 

\subsection{Dependency and Correlation Challenge}

One fundamental challenge in SPS lies in characterizing the dependencies between the social and physical data sources and correlating the collected data across the two sensing paradigms~\citep{stieglitz2018social}. While this challenge has been studied in social and physical sensing independently, it is more pronounced in the context of SPS applications and more challenging to solve due to several hurdles.

The first hurdle is building a unified analytical framework to model the source dependency and data provenance in SPS, given the diversity of source dependencies in social and physical sensing. For example, human sensors tend to be naturally correlated through social networks (e.g., Twitter followers tend to re-tweet their friends' tweets). In contrast, physical sensors do not typically inherit any such social correlations and are more likely to be correlated through the underlying physical phenomena or geographic locations (e.g., two air quality monitors are likely to report similar measurements if they are in close proximity). Such disparity in source dependency and data correlation makes it non-trivial to seamlessly integrate the diverse social and physical sensor measurements under a principled framework \citep{wang2014provenance}. Current knowledge discovery and data mining approaches in social and physical sensing such as semantic pattern recognition~\citep{dey2018fake}, trust and influence modeling~\citep{asim2019trust}, and covariance intersection~\citep{ahn2011new} model the dependencies across social and physical sources independently. However, due to the different source dependencies within the social and physical sensors, such approaches are largely inadequate for SPS applications. A unified source dependency modeling framework to meld the social and physical sensors in SPS is yet to be developed.

\begin{figure}[!ht]
    \centering
    \includegraphics[width=8.5cm]{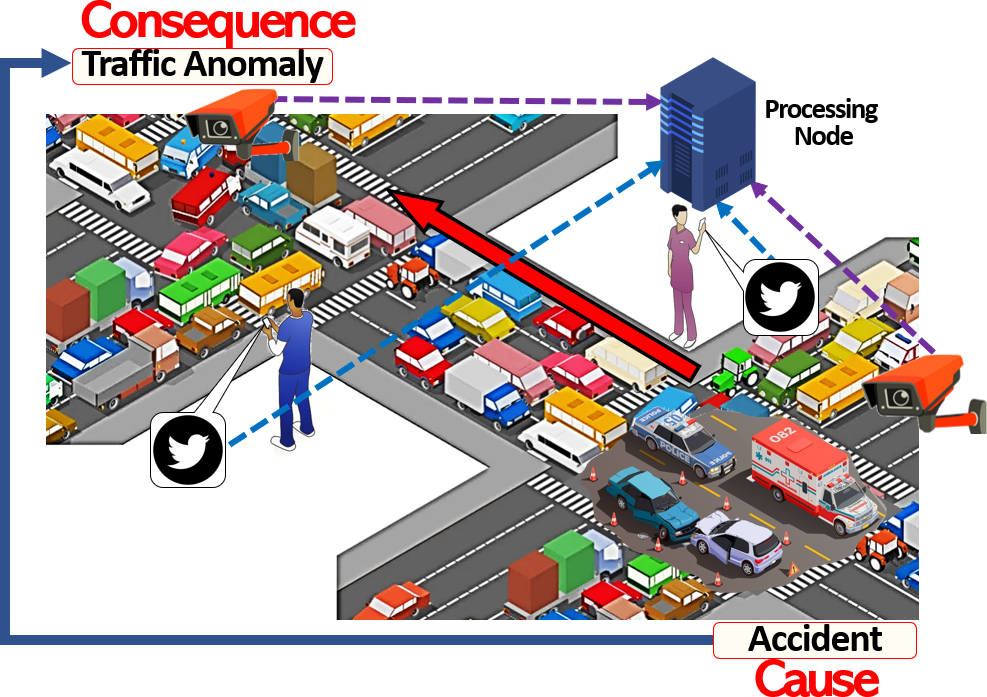}
    \caption{Example of causality among sensors in SPS}
    \label{fig:causality}
\end{figure}

The second hurdle is imposed by the presence of strong causal relationships 
between the physical and social sensor data in SPS applications~\citep{giridhar2016clarisense+}. For instance, during a traffic accident, as illustrated in Figure~\ref{fig:causality}, people might report the accident along with its location on Twitter, while traffic flow monitoring units placed at a different segment on the same road might detect unusually slow traffic movements~\citep{giridhar2016clarisense+}. While the traffic accident and congestion reported by different sensing channels might be seemingly unrelated at first glance, aligning the temporal and spatial information from the input signals (e.g., geolocation information and timestamps of the events) might reveal an inherent causality between them (i.e., the traffic congestion was probably caused due to the traffic accident)~\citep{tsapeli2017non}. Thus, even though there might not be any direct relation between the reported events across social and physical data platforms, the sensors across the two paradigms might report the same chain of occurrences or the same context but in different formats. While this context information might help to explain the cause of anomalous incidents, it is a non-trivial task to explore such causality across social and physical data platforms.

Given the diverse source dependency profiles and the potential presence of causality across the physical and social sensors in SPS, it is challenging to design a holistic framework that can effectively connect the disparate social and physical sensors for interpreting real-world event occurrences. Consequently, extensive exploration and modeling of the dependency and correlation within the social and physical domains remain an outstanding challenge in SPS research.


\subsection{Social and Physical Privacy Challenge}
\begin{figure}[!ht]
    \centering
    \includegraphics[width=8.5cm]{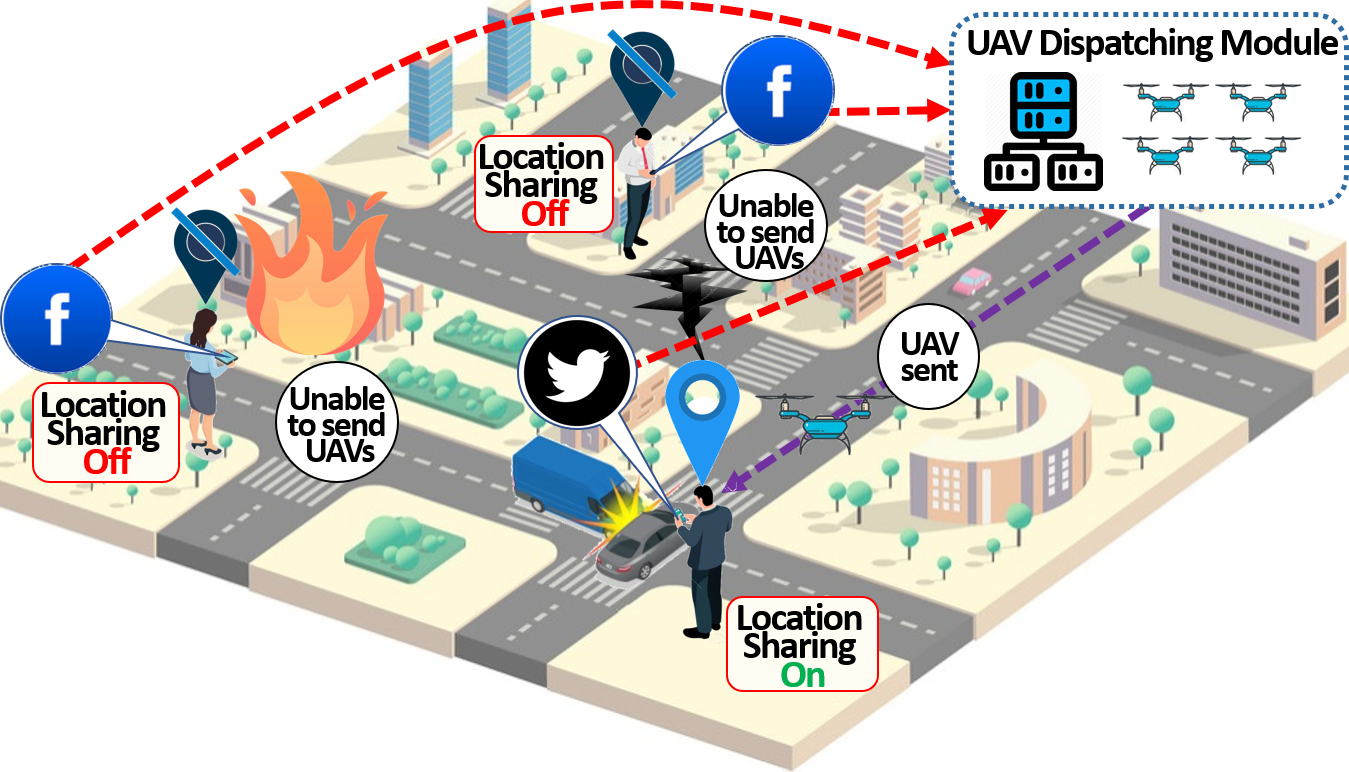}
    \vspace{-0.05in}
    \caption{Example of social and physical privacy challenge in SPS}
    \vspace{-0.15in}
    \label{fig:privacy}
\end{figure}

Due to the integrated nature of the social and physical sensors in SPS, one critical challenge in SPS applications is to efficiently address the privacy issues of end users of SPS applications~\citep{liu2019privacy}. Figure~\ref{fig:privacy} illustrates an SAS application where UAVs need to be dispatched based on locations derived from social media~\citep{terzi2020towards}. However, due to concerns about privacy, a good proportion of users refrain from sharing their GPS data, due to which the UAVs would be unable to determine the locations where to fly to. Existing literature has proposed several privacy-aware sensing approaches for social sensing, which include source identity obfuscation~\citep{toch2012personalization}, blind signatures and data shuffling~\citep{liu2019privacy}, ring signatures~\citep{vance2018privacy}, and data perturbation~\citep{ganti2008poolview}. In a similar fashion, for alleviating privacy issues in physical sensing, current approaches have developed schemes such as slice-mixed aggregation~\citep{li2009privacy}, isolated virtual networks~\citep{al2015internet}, trace-free location tracking~\citep{toch2012personalization}, and routing with random walk~\citep{li2009privacy}. Despite the effectiveness of the above approaches in preserving user privacy in social and physical sensing separately, several unique difficulties in SPS restrict their usefulness in solving the privacy challenge in SPS systems. 

First, social and physical sensors in a connected environment often deliver complementary information that can be exploited to expose the users' personal information. For example, in a fitness tracker application using social media and wearable sensors, reports of daily exercise activities posted by people through social media (e.g., jogging in a park) might be correlated with user-shared historical health data from wearable sensors (e.g., blood pressure, pulse rate, body temperature) to potentially infer the medical history of an individual (e.g., whether a person has a chronic illness).

Conversely, in SPS applications, the data from physical sensors might also be exploited to maliciously extract the private information of individuals when augmented with social signals. For example, in an anomalous crowd detection application that combines images captured by surveillance cameras with reports of crowd gatherings posted on social media to infer the onset of sudden crowds, the surveillance cameras can only capture the image of a person at a specific location without further details of that person. However, if that particular person periodically shares their shopping history alongside geo-location information on social media, the image data from the cameras might be correlated with the additional data to unravel the socioeconomic status of the individual~\citep{xiong2019personalized}.

Second, due to the inherent heterogeneity of the devices and data in SPS, it is a non-trivial task to apply unified privacy-preserving policies in SPS applications. As discussed in the \textit{device and data heterogeneity challenge}, SPS applications involve diverse devices. With such a wide range of devices, it is difficult to keep track of the data transmission and security protocols of all the devices. As such, device vulnerabilities such as unprotected APIs, outdated firmware, or defunct authentication mechanisms~\citep{hasan2016development} might be exploited by hackers to steal personal data from user devices. Moreover, since social and physical sensors in SPS generate a wide variety of data (e.g., text, image, video, audio, location data), the capture, transmission, and processing of the data require different energy profiles, which often leaves the devices in SPS vulnerable to exploits such as a side-channel attack, an attack intended to steal user data~\citep{lerman2011side}. For example, when a device is processing video frames, the patterns of power usage within the device might be analyzed by an attacker to recover the raw video data~\citep{abrishamchi2017side}. Current privacy-preserving approaches are not designed to withstand the intrinsic and pronounced data and device heterogeneity prevalent in SPS applications, which might lead to vulnerability of user privacy. Thus, it is yet to be determined how to design unified privacy-aware SPS platforms that can concurrently consider the data and device heterogeneity and protect sensitive user information to address the privacy challenges in SPS.

\subsection{Interrelated Dynamics Challenge}
A pivotal challenge in SPS is handling the interrelated dynamics induced by the fusion of the social and physical domains. SPS applications innately rely upon the tight integration between social and physical realms, both of which are dynamic in nature and exert impact over one another. 

\begin{figure}[!ht]
    \centering
    \includegraphics[width=8.5cm]{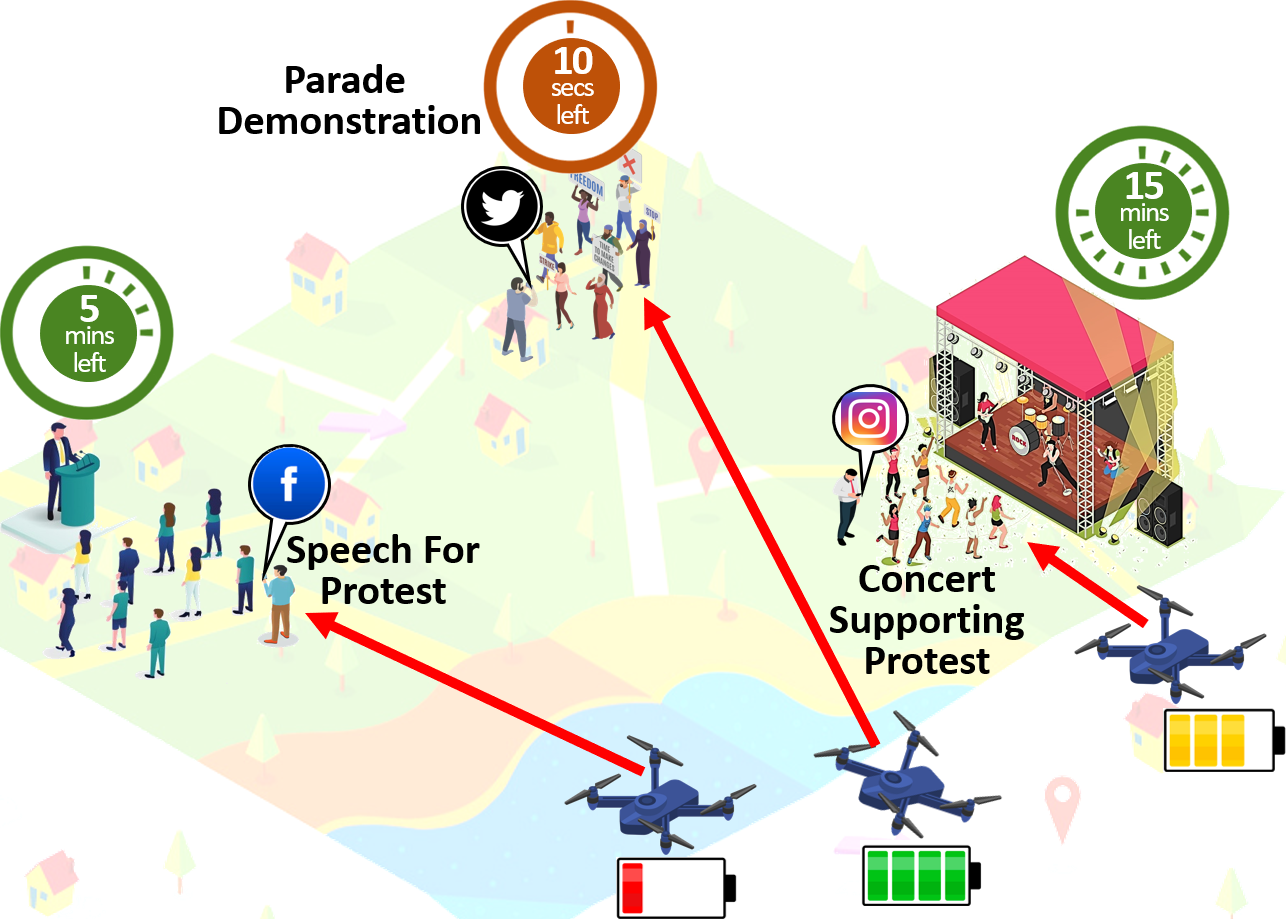}
    \vspace{-0.05in}
    \caption{Example of how social domain dynamics affect physical sensors in SPS}
    \label{fig:dynamics1}
\end{figure}

The dynamics arising from the social domain tend to influence the performance of physical sensing directly. Let us consider an integrated social media and UAV-driven crowd analysis application as shown in Figure~\ref{fig:dynamics1}~\citep{kaiser2017advances}. If social events related to public protests are initiated and organized on social media, dynamics in the social domain (e.g., more people tweeting, different locations being targeted, people publicizing the activities to a greater level) might cause dynamics in the physical world (e.g., more new activities related to protests such as speeches and concerts, more people joining, events taking place in locations far away from one another). Given that mobile physical sensors such as UAVs and robots often suffer from constraints such as energy, communication, and speed, such physical sensors might not be able to explore or investigate all the events reported by social sensors within set deadlines. Such a scenario is also illustrated in Figure~\ref{fig:dynamics1}, where the initiated crowd events are located at various locations with different deadlines. Due to the presence of the social domain dynamics, the UAVs, with their physical constraints, might not be able to sense all the crowd events before their deadlines. Thus, careful choices need to be made on which subset of reports from the social data platforms to prioritize for the physical sensors, which existing solutions have not addressed.

\begin{figure}[hbt!]
    \centering
    \includegraphics[width=8.65cm]{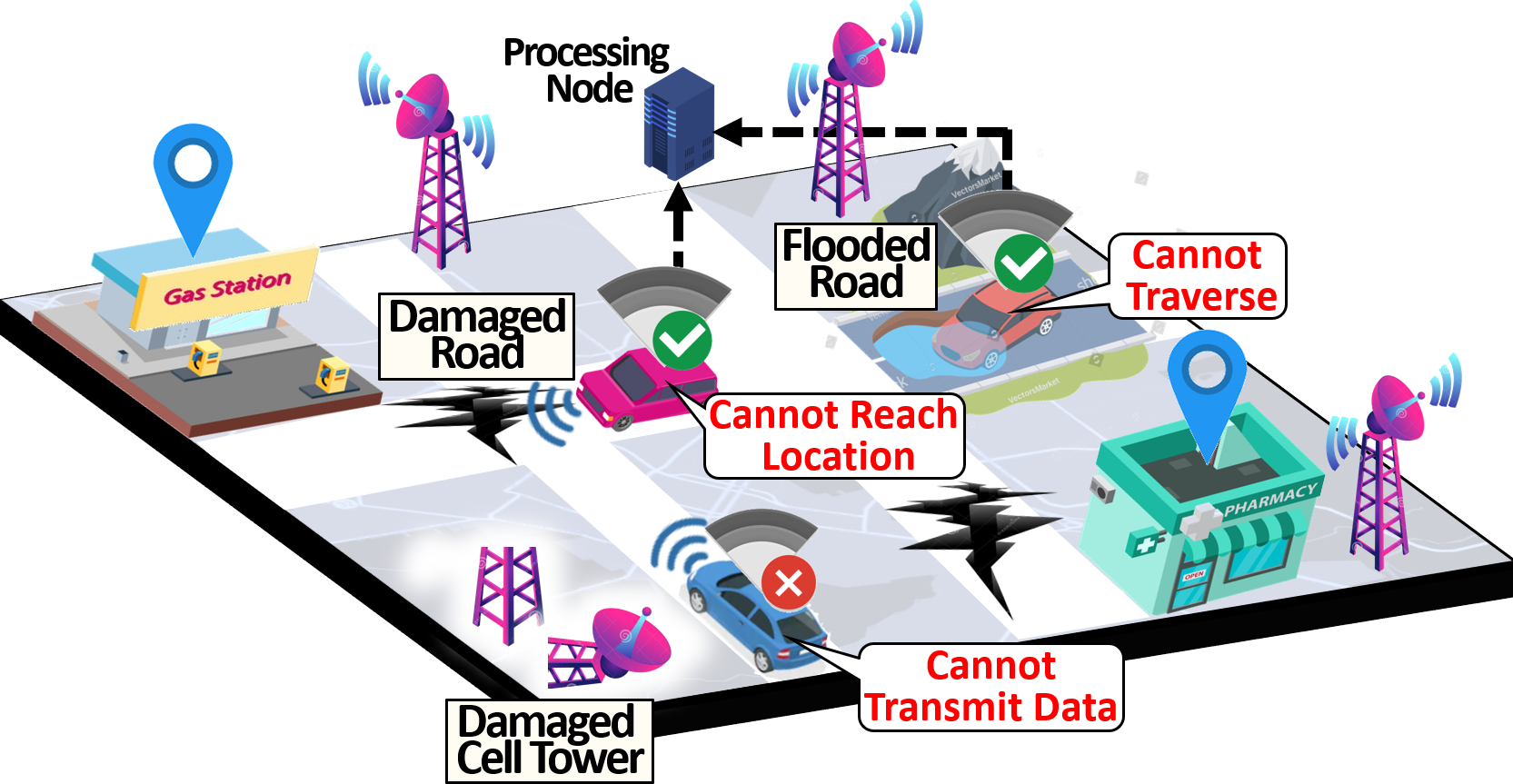}
    \vspace{-0.05in}
    \caption{Example of how physical domain dynamics affect social sensors in SPS}
    \vspace{-0.15in}
    \label{fig:dynamics2}
\end{figure}

On the other hand, the dynamics from the physical world might affect the performance of both the social and physical sensors~\citep{hu2012review}. For example, let us consider an S-VSN application in the aftermath of a disaster~\citep{rettore2019vehicular} as shown in Figure~\ref{fig:dynamics2}. The disaster might cause road damage around the affected area. Such damage can potentially restrict the travel of cars across certain roads, which can cause car drivers to be unable to locate and report events of interest on social media (e.g., gas availability in a gas station). Moreover, network infrastructure can also get damaged, leading the car drivers to lose access to network connectivity and be unable to post any event reports~\citep{wisitpongphan2007routing}. Eventually, fewer observations might be reported by car drivers across social media, yielding poor coverage from the human sensors. In the physical world, the event occurrences might also be accompanied by unforeseeable circumstances such as unfavorable weather (e.g., extreme temperatures) or damaged infrastructure (e.g., disconnected power lines), which might impede physical sensing. For example, strong wind or cloud might impact the readings from different sensors such as cameras or gyroscopes on UAVs, and bumpy roads might negate the performance of vehicular sensors on cars (e.g., shaky images captured by dashcams)~\citep{li2019introduction,wang2013exploitation}. Therefore, careful consideration must be given to adapting the SPS systems to accommodate such physical world dynamics on-the-fly, which has not been extensively explored by current literature.

\begin{figure*}[!ht]
    \centering
    \includegraphics[width=18cm]{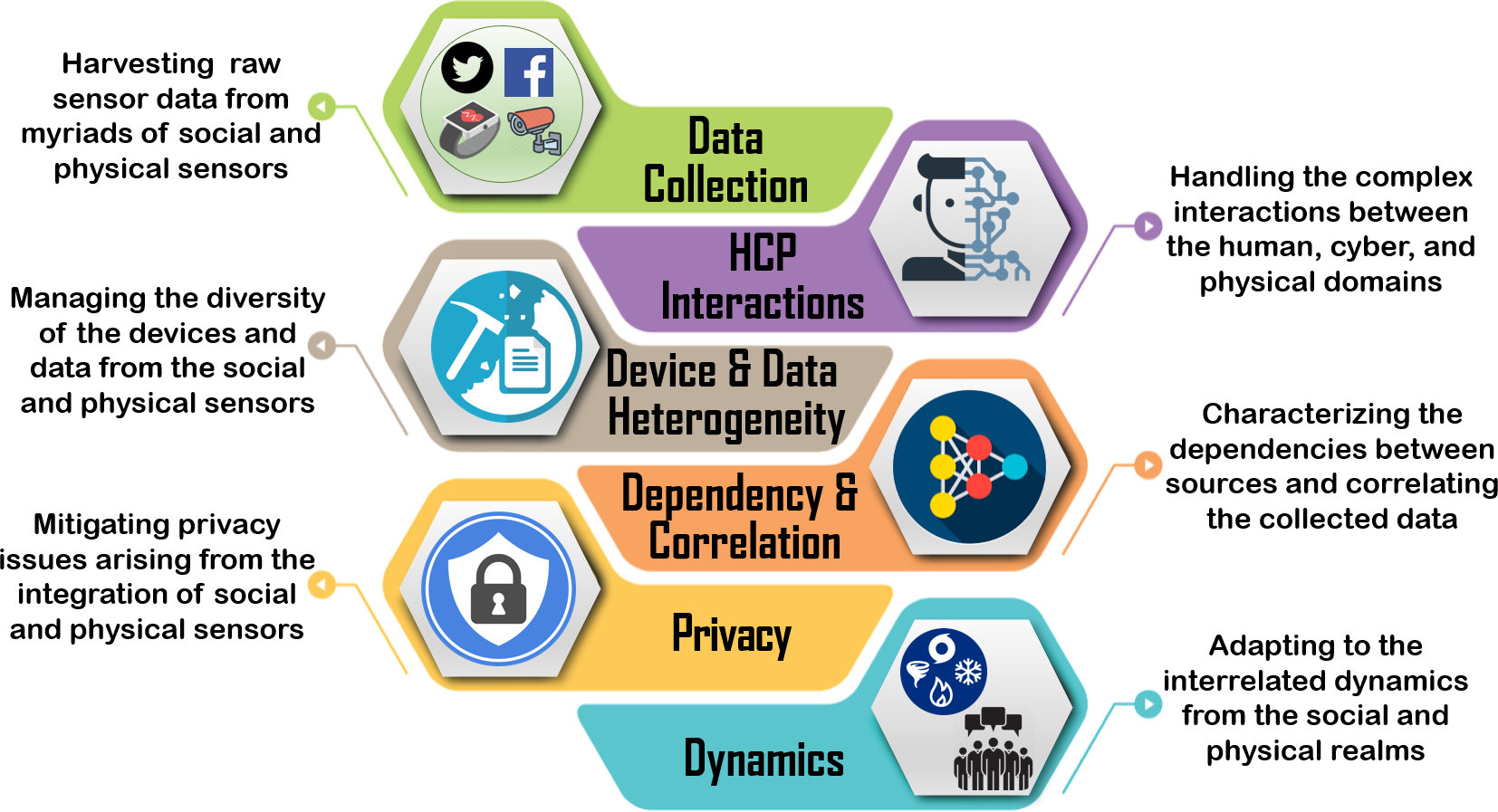}
    \vspace{-0.05in}
    \caption{Overview of fundamental challenges in SPS Applications}
    \vspace{-0.15in}
    \label{fig:Challenges}
\end{figure*}

Figure~\ref{fig:Challenges} provides an overview of the fundamental challenges in SPS. We note that while some of these challenges might also be studied in AI literature, the two areas are sufficiently different and not directly comparable to each other for several reasons. First, SPS is a sensing paradigm that leverages the collective knowledge from human and physical sensors to perceive the state of the world~\citep{qiu2016integrating,de2017cyber}. By contrast, AI is a much broader topic that encompasses the theories and algorithms driving systems that can perform tasks that typically require human intelligence~\citep{joiner2018emerging}. Second, SPS and AI have fundamentally dissimilar problem contexts. For instance, while \textit{data heterogeneity} is also studied in AI literature, for SPS, the problem context is unique because: (a) SPS applications are often involved with a diversity of tasks (i.e., data capture, communication, and computation) and sensing sources (e.g., social media, UAVs, crowdsourcing participants)~\citep{zhang2019edgebatch}, which are less likely to be present in general AI applications; (b) the entities in SPS (i.e., humans and machines) often have a complex interdependence of workflow~\citep{zhu2019delay}, a problem which is not frequently encountered in AI applications (Wei et al. 2019); and (c) the social and physical sensors in SPS are known to produce data at different frequencies which injects an inherent complexity in the data acquisition process~\citep{misra2020iot, mourtzis2016industrial}. In a similar fashion, the \textit{dependency and correlation} challenge is also studied by AI literature, but solutions designed for AI applications might be inapplicable to SPS applications because: (a) the source dependencies within the social and physical sensors in SPS are inherently different and; (b) the data from physical and social sensors in SPS applications often have strong latent causal relationships that are not explicitly considered in general AI solutions~\citep{giridhar2016clarisense+}. In the following section, we explore possible directions for future research to address the above challenges in SPS.

	\section{Roadmap for Future Work} \label{sec:directions}
In this section, we present several exciting avenues for future work in the domain of SPS. As we outline each avenue, we enlist a few potential directions of research to pursue. 

\subsection{Uncertainty Quantification in SPS}
Since SPS applications often rely on the noisy social and physical signals contributed by a diversified set of human and physical sensors, one potential direction for future work lies in quantifying the uncertainty generated by the diverse sensors in SPS applications. As discussed in the \textit{data collection} challenge in Section~\ref{sec:challenges}, the intrinsic interdependence between the noise generated by the social and physical sensors in SPS is hard to quantify and model. As such, the collected social and physical signals induce a degree of uncertainty. Without carefully determining the level of uncertainty in the input data, the performance of SPS applications might be unpredictable~\citep{wang2017service,zhang2018scalable,zhang2017constraint}. Current social sensing analytics tools such as \textit{truth discovery} algorithms primarily focus on deducing the data veracity or source reliability from the social sensing data \citep{ali2011social,huang2016topic,zhang2018towards}. In a similar fashion, current physical sensor data processing schemes have focused on inferring the information contained in the physical signals using techniques such as fuzzy logic~\citep{ma2006fuzzy}, autoregressive models~\citep{cockx2014quantifying}, arbitrary polynomial chaos~\citep{gulgec2020uncertainty}, perturbation theory~\citep{zhao2014quantifying}, and full factorial numerical integration~\citep{lee2009comparative}. However, the existing schemes do not explicitly focus on quantifying the interconnected uncertainty between the social and physical sensors, which is important to ensure the stable performance of SPS systems. As illustrated in Figure~\ref{fig:datacollect} in Section~\ref{sec:challenges}-A with the example of a structural damage assessment application, unreliable online users might incorrectly report the damage sites, which might dispatch robots to the wrong locations. Likewise, if the images captured by the robots are obscure, they might not correctly validate the reliability of the users. Thus, the characteristic noise originating from social and physical sensors might adversely affect the sensing performance of the SPS applications. As such, it is imperative to develop methods for quantifying the uncertainty signals in SPS applications. 

It is essential first to realize why existing literature has not extensively explored the domain of uncertainty quantification in SPS. Several disparities between social and physical sensors lead to difficulty in rigorously quantifying their signals' uncertainty. First, the social and physical sensors in SPS generate dissimilar types of data (e.g., social sensors typically generate text data while physical sensors generate continuous and discrete time signals)~\citep{mitchell2014survey,wang2019age}. Second, the dependencies between the sensors in social data platforms are different from that within the sensors in physical data platforms~\citep{stieglitz2018social}. Third, the dynamics in the social domain are characteristically contrasting to the dynamics in the physical world~\citep{hu2012review,zeng2020survey}. Fourth, the rates of data generated by social and physical sensors are different from physical sensors (e.g., the speed at which UAVs capture images is different from the frequency at which people report incidents on Twitter)~\citep{misra2020iot,mourtzis2016industrial}. In addition, factors such as biased opinions from human sensors in social sensing and the failure cases of physical sensors (e.g., out of battery or affected by bad weather) implicitly aggravate the uncertainty quantification in SPS~\citep{wang2019age,diez2020local}.

One direction for further research in SPS is to focus on rigorously quantifying the uncertainty of social and physical signals and leverage the quantification results to improve SPS systems' social and physical sensing components jointly. For example, in anomaly detection with an SAS application, if the uncertainty from the social signals can be determined, it may help to dispatch the UAVs better. Similarly, if the uncertainty in the captured UAV data can be measured, it can be used as feedback signals to improve reliable source selection in social sensing. Another probable research direction in SPS can be to design schemes that can deduce the uncertainty in the social and physical sensing data while simultaneously considering the SPS challenges such as the data heterogeneity, the diverse source dependencies, the social and physical world dynamics, and the contrasting social and physical sensor data generation rates. Existing studies on statistical analysis have proposed principled approaches based on estimation theory. Examples of uncertainty quantification approaches include maximum likelihood estimation (MLE), Cramer-Rao lower bounds (CRLB)~\citep{wang2013credibility,wang2011quantifying,wang2011bayesian,wang2012scalability,wang2015reliable}. Alongside quantifying the uncertainty of estimation results, future SPS schemes can focus on incorporating the accompanying factors (e.g., human bias, physical constraints) in the uncertainty propagation models. 
We envision that techniques from multiple disciplines might be applicable for alleviating the above hurdles and modeling the uncertainty in SPS applications which includes Bayesian networks~\citep{zhang2018streamguard}, Monte Carlo methods~\citep{harris2014monte},  evidence theory~\citep{bae2004approximation}, Markov Chain formulation~\citep{abdar2020review}, mixed integer linear programming (MILP)~\citep{constantinescu2010computational}, and polynomial chaos expansion~\citep{kaintura2018review}.

\subsection{Handling Trade-Off Between Privacy and Sensing Quality in SPS}
\begin{figure}[!ht]
     \centering
     \includegraphics[width=8.5cm]{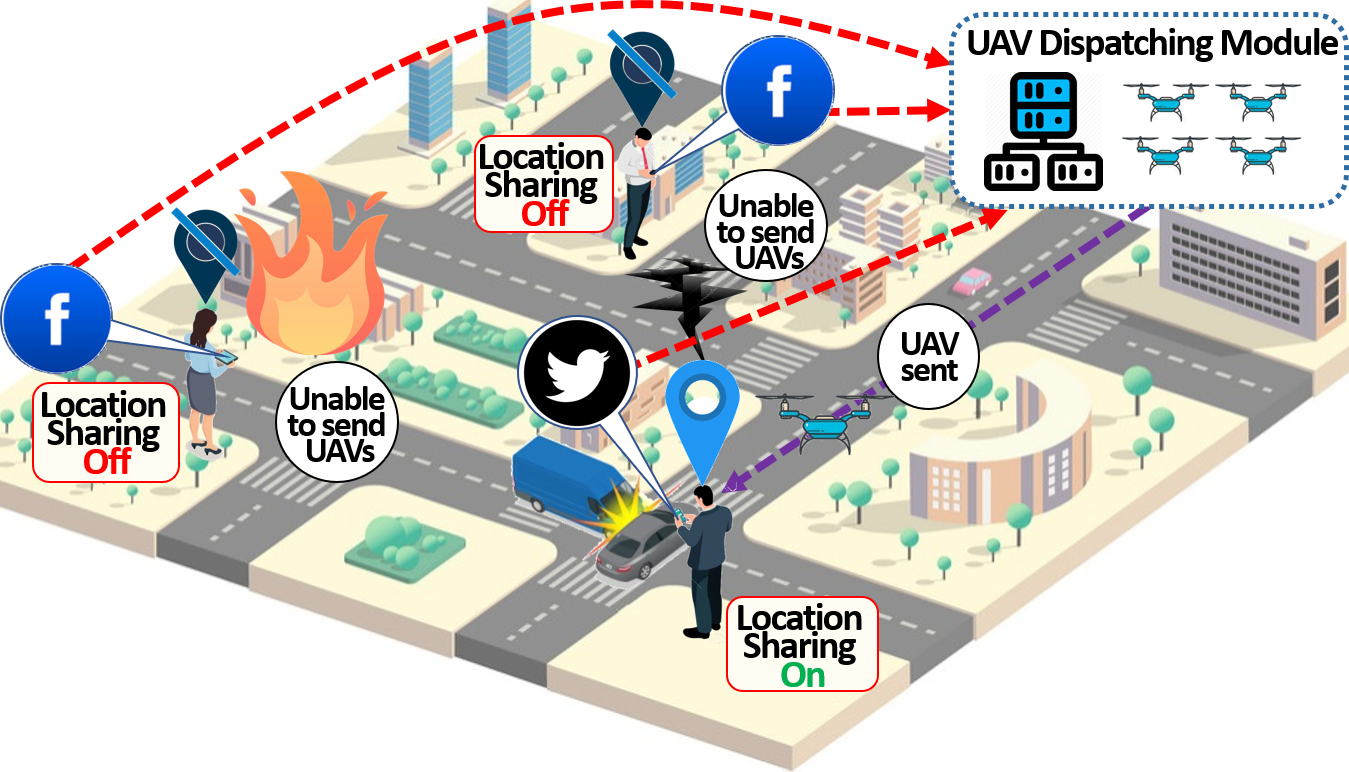}
     \vspace{-0.05in}
     \caption{Example of trading-off between privacy and sensing quality in SPS}
     \label{fig:tradeoff}
\end{figure}

As identified in Section~\ref{sec:challenges}-E, mitigating privacy issues is critical in SPS applications. However, in attempts to ensure user privacy, often current SPS schemes have to compromise the sensing quality. For example, metadata such as geo-location information might be concealed from privately-owned devices to protect the identity of human sensors on social media. However, the location information might be critical for mobile sensors, such as robots, to be dispatched to events of interest. Figure~\ref{fig:tradeoff} shows an example scenario where concealing private data affects sensing quality. Thus, SPS applications often require the knowledge of supporting information such as locations, timestamps, and contextual information from reported social sensing data, which often conflicts with the end users' privacy requirements. Since ensuring user-level privacy and maximizing sensing quality often turn out to be two potentially conflicting objectives in SPS~\citep{xu2018conflict}, it is imperative to design schemes that carefully strike trade-offs between privacy and sensing quality for an optimized SPS system.

Existing approaches in data-driven social and physical sensing schemes have proposed techniques to manage user privacy by obfuscating identifying information such as geo-location tags from the raw data from personal devices (e.g., laptops, smartphones)~\citep{park2017protecting,zhang2018privacy}. However, existing privacy-preserving schemes have not addressed effectively handling the trade-off between privacy and sensing quality in SPS. Several reasons make it difficult to simultaneously establish privacy and sensing quality in SPS. First, the unpredictable nature of human users in SPS applications makes it difficult to ensure that the users will strictly abide by policies to protect their privacy. Second, given the unique data and device heterogeneity in SPS applications, designing a unified framework to enforce individual privacy policies across all the devices is a challenging task~\citep{vance2019towards}. Third, regardless of the robustness of privacy-preserving schemes, the complementary aspects of the contributed data through social and physical data platforms can be exploited to steal sensitive user information~\citep{vance2018privacy}.

One future research direction to pursue for optimizing the privacy and sensing quality in SPS applications is to design multi-faceted cryptographic techniques such as blockchain technology~\citep{ali2017iot}, smart contracts~\citep{christidis2016blockchains}, and ring signatures~\citep{vance2018privacy}. While existing cryptographic approaches have come a long way in balancing privacy and sensing quality individually in social and physical sensing~\citep{henry2018blockchain}, it is difficult to apply unified cryptography-based solutions in an SPS setting where a diverse range of devices are associated~\citep{marin2015optimized}. 
Future work in this domain can constitute developing cryptographic SPS approaches that can cater to the heterogeneity of the devices in SPS and effectively trade off privacy and sensing quality. Another potential direction for future work on quality-aware privacy preservation in SPS is to explore and incorporate approaches like differential privacy, where noise is deliberately added to the user data to conceal the sensitive information of users~\citep{abadi2016deep,kairouz2015composition}. While current differential privacy techniques have been applied in participatory sensing and crowdsensing, such approaches have not considered the data heterogeneity issue prevalent in SPS. Due to the wide variety of the data generated by the social and physical sensors in SPS (e.g., text, image, audio, and location data), injecting deliberate noise for concealing user identity into different data might be computationally intensive and resource-demanding. As such, further work can concentrate on alleviating the data heterogeneity in SPS applications by developing efficient differential privacy techniques.

\subsection{Ensuring Fairness and Accuracy of Detection in SPS}
\begin{figure}[!ht]
     \centering
     \includegraphics[width=8.5cm]{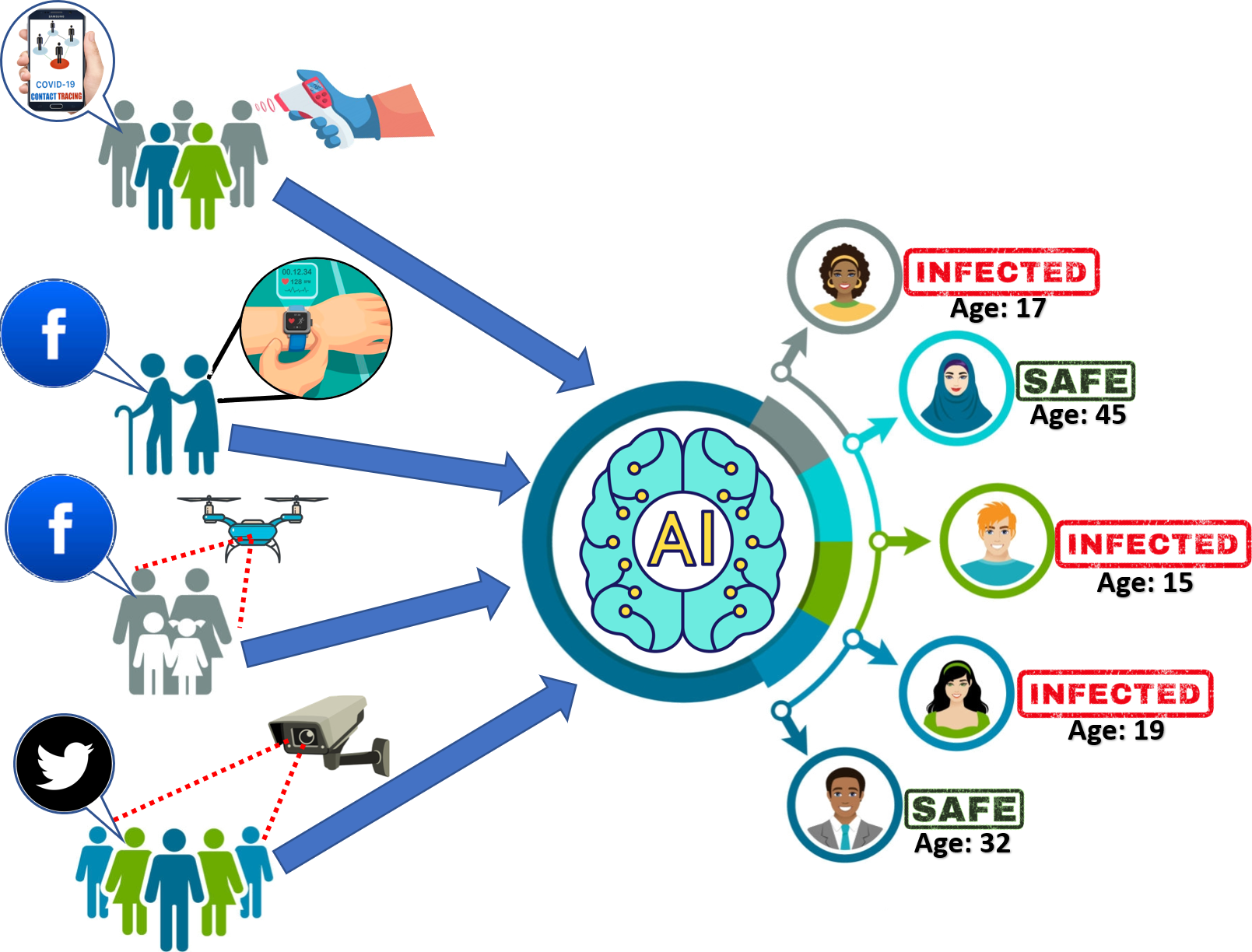}
     \caption{Example of algorithmic bias affecting fairness in an SPS-based contact tracing application}
     \label{fig:fairness}
 \end{figure}

While SPS applications deliver a multifaceted sensing package using a combination of social and physical sensors, one remaining issue is ensuring fairness alongside accuracy for the data obtained from diverse demographics~\citep{kairouz2019advances,zhang2020fairfl}. With the advent of numerous data acquisition platforms and processing techniques, there is a heightening concern from various civil rights organizations, governments, and analysts regarding the fairness of the detection process in SPS applications and their prevalent algorithmic bias towards specific demographic groups~\citep{roselli2019managing}. For example, in a contact tracing SPS application, as illustrated in Figure~\ref{fig:fairness}, overrepresented classes of data might cause a certain age of people (e.g., teenagers) to be incorrectly represented as the prime sources of the disease. One issue that arises when trying to ensure fairness and accuracy in input data distribution is the loss of model accuracy. Specifically, in order to reduce the bias, it is crucial to incorporate a wider distribution of data from different classes (e.g., race, ethnicity, nationality) of data contributors. However, to reduce bias by incorporating a wider distribution of data, the inference models in SPS need to train over a larger sample of data, causing the overfitting/underfitting problem, which often leads to the reduction in model accuracy~\citep{dressel2018accuracy}. The fairness and accuracy issue in SPS is further exacerbated by the fact that specific demographics might be more inclined to use smart devices more often than others. For example, in an anomalistic crowd investigation application using SAS, younger people might use their mobile devices to post crowd-related events more frequently on social media while senior people might not report their observations so often on social media. As such, a crowd inference model might be overfitted with a younger demographic.
Current fairness and accuracy optimizing schemes are limited in addressing such diverse device usage scenarios present in SPS applications.

Existing schemes have attempted to reduce algorithmic bias by using heuristic approaches such as genetic algorithm~\citep{kosmidis2010generic}, optimizing the model's loss function~\citep{iosifidis2019adafair}, or ensuring that the model training process satisfies the given fairness constraints~\citep{zhang2019faht}. The problem with current approaches is that they have been originally designed for fairly good-quality input data. However, in the context of SPS applications, both social and physical sensors are prone to systematic noise, which is hard to quantify and model due to their complex interdependence with each other~\citep{qu2020gan}. Thus, further research can concentrate on optimizing the fairness and accuracy of SPS applications while concurrently offsetting the noise generated by the social and physical sensors. Techniques such as deep learning (DL)-based collaborative filtering~\citep{bobadilla2020deepfair}, discrimination-aware channel pruning~\citep{zhuang2018discrimination}, and selective adversarial networks~\citep{adel2019one} could be explored to develop such fairness and accuracy-optimizing methods. In addition to mitigating the noise contained in the input signals in SPS, one strand of research can be focused on developing user-friendly and accessible interactive interfaces (e.g., interactive kiosks, smartphone applications, responsive websites, augmented reality experiences) for collecting fair data samples in SPS given the potential demographic bias in the participants.

\subsection{Harnessing Adaptive Artificial Intelligence (AI) in SPS}
One route for future research in SPS can be focused on addressing the interrelated dynamics in SPS. As discussed in Section V-F, a critical task in SPS applications is handling the interrelated dynamics caused by the constantly transitioning social and physical environments. Adaptive Artificial Intelligence (AI) algorithms are known to adjust their parameters to cater to changing stimuli~\citep{mcmahan2021advances}. As such, AI algorithms might help to adapt to the constantly changing social and physical environments. However, several limitations inhibit off-the-shelf AI algorithms from being directly applied to SPS applications to mitigate the dynamics challenge. 

First, as identified in Section~\ref{sec:challenges}-B, one recurring issue stemming from the human-cyber-physical interactions challenge in SPS applications is the inconsistent availability of the social and physical sensors, known as \textit{churn}~\citep{vance2019towards}. Many AI algorithms heavily rely on the sensing devices' participation in the training phase, which requires multiple iterations to converge to global optima. Given the churn involved in SPS applications, it is often difficult for AI algorithms to classify the incoming sensing measurements accurately. As a result, these AI algorithms might end up with failure scenarios in SPS applications with significant dynamics. 
Second, SPS applications typically involve a large number of privately-owned devices, and often users do not provide access to their devices with concerns about privacy or excessive bandwidth usage. Traditional distributed AI algorithms often tend to assume unrestricted access to local datasets from individuals' devices~\citep{mcmahan2021advances}, which may not always hold in SPS applications. 
Given the inaccessibility of user data across privately-owned devices, existing distributed AI algorithms fall short of addressing the dynamics challenge in SPS applications. Third, SPS applications involve a diverse set of devices and a wide range of data types (i.e., data and device heterogeneity). However, most current AI algorithms are intended to handle input data that is naturally homogeneous and assume that the data is identically distributed across the devices~\citep{li2020federated}. While a few existing distributed AI algorithms can handle heterogeneous data, the computational complexity of such algorithms tends to be relatively high, which might overload resource-constrained devices (e.g., smartphones, UAVs) used in SPS applications~\citep{mcmahan2021advances}. Consequently, such limitations make it challenging to apply existing distributed AI algorithms 
to critical SPS applications.

Several future avenues for research can be explored to tackle the above difficulties. One potential realm of further work can focus on using \textit{deviceless pipelining} techniques to offload and distribute AI model training subtasks in SPS applications across devices equipped with specialized hardware~\citep{vance2019towards}. For example, in a disaster response application with SAS, a UAV fitted with a GPU having large video RAM can be used to execute grid search for hyperparameters in AI model training while another UAV fitted with an FPGA can be used for pooling and flattening subtasks. A second emerging model training technique is to incorporate human intelligence (HI) for augmenting AI algorithms and enhancing their performance~\citep{amershi2019guidelines}. HI platforms such as Amazon MTurk have allowed human participants to provide their inputs for labels or features that might be potentially leveraged to retrain the AI models and address their innate flaws~\citep{zhang2019crowdlearn}. 
Thus, further research in SPS can focus on incorporating HI with AI to develop robust human-AI algorithms for SPS applications. A third probable future avenue of research can focus on designing decentralized model training algorithms for collaboratively acquiring local model updates from privately-owned devices. With the intent of preserving privacy and reducing network bandwidth requirements, federated learning (FL) is gaining traction as a decentralized AI training paradigm~\citep{konevcny2016federated,zhang2021fedsens}, where a shared global AI model is trained from a collection of edge devices owned by end users~\citep{wang2019social}. Future research can focus on developing FL solutions that can consider the data and device heterogeneity originating from the social and physical sensors in SPS.

	\section{Conclusion}\label{sec:conclusion}
In this paper, we present a comprehensive survey of SPS, an emerging integrated sensing paradigm that exploits the collective strengths of physical and social sensing to acquire and interpret observations from the environment. Empowered by the ubiquity of versatile data capture, communication, and computing technologies, SPS melds the human wisdom-driven data acquisition from social sensors with the multifaceted sensing capabilities of physical sensors to deliver a deeper perception of the real world, both physically and socially. In particular, this paper surveys the various aspects that are important for constructing compelling SPS systems, which includes a detailed overview of SPS, the key motivation behind its origin, the crucial technologies and protocols that enable SPS, real-world SPS applications and state-of-the-art solutions, the key challenges prevalent in SPS, and the potential avenues for further work to address the challenges. We hope this paper will bridge the knowledge gap from the current literature on SPS and motivate future studies to design novel SPS systems for a more holistic perception of real-world phenomena.

\section*{Acknowledgment}

This research is supported in part by the National Science Foundation under Grant  IIS-2008228, CNS-1845639, CNS-1831669, Army Research Office under Grant W911NF-17-1-0409. The views and conclusions contained in this document are those of the authors and should not be interpreted as representing the official policies, either expressed or implied, of the Army Research Office or the U.S. Government. The U.S. Government is authorized to reproduce and distribute reprints for Government purposes notwithstanding any copyright notation here on.

\bibliographystyle{SageH}
\bibliography{refs} 

\newpage
\section*{Appendix} \label{sec:appendix}

 \begin{table}[htb!]
     \caption{List of Abbreviations and Acronyms}
     \vspace{-0.1in}
     \center
     \begin{tabular}{|l|l|}
         \hline
         \centering
          AI & artificial intelligence\\ \hline
          ALPR & automatic license plate recognition \\ \hline
          ANNs & artificial neural networks \\ \hline
          API & Application Programming Interface\\ \hline
          CoAP & Constrained Application Protocol  \\ \hline
          COVID-19 & coronavirus disease 2019\\ \hline
          CPS & cyber-physical systems\\ \hline
          CPSS & cyber-physical-social systems\\ \hline
          CRLB& Cramer-Rao lower bounds\\ \hline
          CSS & cyber-social systems\\ \hline
          DDA & disaster damage assessment\\ \hline
          DL & deep learning \\ \hline
          ENS & Exposure Notification System\\ \hline
          FL & federated learning \\ \hline
          GPS & Global Positioning System \\ \hline
          HCI & human computer interactions \\ \hline
          HCP & human, cyber, and physical \\ \hline
          HI & human intelligence \\ \hline 
          IoT & Internet of Things \\ \hline
          LR-WPAN & low-rate wireless personal area network\\ \hline
          LTE & Long-Term Evolution \\ \hline
          MCS & mobile crowdsensing \\ \hline
          MDP & Markov Decision Process\\ \hline
          MEC & mobile edge computing \\ \hline
          MILP & mixed integer linear programming \\ \hline
          ML &  machine learning\\ \hline
          MLE & maximum likelihood estimation\\ \hline
          MQTT & Message Queue Telemetry Transport\\ \hline
          PID & proportional–integral–derivative \\ \hline
          RFID & radio frequency identification \\ \hline
          RSSI & received signal strength indicator\\ \hline 
          RSU & roadside units \\ \hline
          SAS & social airborne sensing\\ \hline
          SPS & social-physical sensing\\ \hline
          S-VSN & social vehicular sensor network\\ \hline
          UAV &  unmanned aerial vehicles,\\ \hline
          UDP & User Datagram Protocol \\ \hline
          UGV &  unmanned ground vehicles,\\ \hline
          VR & virtual reality\\ \hline
          VSN & vehicular sensor networks\\ \hline
          WSN & wireless sensor networks\\ \hline
          XMPP & Extensible Messaging and Presence Protocol\\ \hline
     \end{tabular}
     \label{tab:definition}
 \end{table}

\end{document}